\documentclass[journal,comsoc]{IEEEtran}

\usepackage{xspace,amsmath,amssymb,amsfonts,epsfig,syntonly}
\usepackage{cite,bm,color,url,textcomp,empheq,boxedminipage}
\usepackage{algorithmicx,algorithm,algpseudocode}
\usepackage{epstopdf}
\usepackage{empheq}
\usepackage{pifont}
\usepackage{setspace}

\usepackage{graphicx,graphics,framed,subfigure}  
\usepackage{multirow,multicol}
\usepackage{psfrag}    
\usepackage{stfloats}
\usepackage{url}

\newtheorem{definition}{\underline{Definition}}[section]

\newtheorem{lemma}{{Lemma}}

\newtheorem{example}{{Example}}

\DeclareMathOperator*{\argmax}{arg\,max}
\DeclareMathOperator*{\argmin}{arg\,min}

\long\def\symbolfootnote[#1]#2{\begingroup
\def\thefootnote{\fnsymbol{footnote}}
\footnote[#1]{#2}\endgroup}

\psfull

\allowdisplaybreaks[4]

\usepackage[T1]{fontenc}
\usepackage{ifpdf}
 \ifpdf
 \else
 \fi

\begin{document}

\title{Dynamic RAT Selection and Transceiver Optimization for Mobile Edge Computing Over Multi-RAT Heterogeneous Networks}

\author{Feng~Wang,~\IEEEmembership{Member,~IEEE,}~and~Vincent~K.~N.~Lau,~\IEEEmembership{Fellow,~IEEE}
\thanks{F. Wang is with the School of Information Engineering, Guangdong University of Technology, Guangzhou 510006, China, and is also with the Department of Electronic and Computer Engineering, The Hong Kong University of Science and Technology, Hong Kong (e-mail: fengwang13@gdut.edu.cn)}
\thanks{V. K. N. Lau is with the Department of Electronic and Computer Engineering, The Hong Kong University of Science and Technology, Hong Kong (e-mail: eeknlau@ust.hk).}
}

\maketitle

\begin{abstract}
  Mobile edge computing (MEC) integrated with multiple radio access technologies (RATs) is a promising technique for satisfying the growing low-latency computation demand of emerging intelligent internet of things (IoT) applications. Under the distributed MapReduce framework, this paper investigates the joint RAT selection and transceiver design for over-the-air (OTA) aggregation of intermediate values (IVAs) in wireless multiuser MEC systems, while taking into account the energy budget constraint for the local computing and IVA transmission per wireless device (WD). We aim to minimize the weighted sum of the computation mean squared error (MSE) of the aggregated IVA at the RAT receivers, the WDs' IVA transmission cost, and the associated transmission time delay, which is a mixed-integer and nonconvex problem. Based on the Lagrange duality method and primal decomposition, we develop a low-complexity algorithm by solving the WDs' RAT selection problem, the WDs' transmit coefficients optimization problem, and the aggregation beamforming problem. Extensive numerical results are provided to demonstrate the effectiveness and merit of our proposed algorithm as compared with other existing schemes.
\end{abstract}


\section{Introduction}
 Mobile edge computing (MEC) has been attracting significant research interest over the past few years\cite{Sar14,Chiang16,JunZhang17}. By distributing computation-intensive tasks to the edge cloud server, mobile devices can collaboratively handle low-latency applications (such as augmented/virtual reality (AR/VR) and online gaming) which require huge computational and storage capacity. In fact, the success and performance of MEC depend heavily on the computational model of tasks. The earlier works\cite{Feng18,TCOM19,Bi2020,JunZhang17} considered several simplified computational models, which were very far from the actual tasks.

 Recently, the MapReduce framework has been proposed as a realistic model for distributed computing \cite{MapReduce-08,MapReduce-10}. In MapReduce framework, a large dataset is split into multiple data chunks and stored distributively across multiple wireless devices (WDs). Then Map, Shuffle, and Reduce phases are implemented to compute several task functions\cite{MapReduce-10}. Specifically, in the Map phase, each map task of the WDs locally reads one data chunk and generates an intermediate value (IVA) in parallel. These IVAs are then exchanged by the WDs via the wireless network in the Shuffle phase. Finally, in the Reduce phase, the WDs fetch the IVAs from the assigned subset of the dataset and apply the Reduce function to produce the final results. However, such uploading of IVAs in the Shuffle phase will consume valuable communication resources in mobile networks, and it is critical to determine which IVAs to upload and when to upload them to strike a balance between computational performance and communication overhead.

 In the literature, there have been some works on resource optimization in MapReduce to improve communication efficiency in the Shuffle phase. For example, the works \cite{SongzeLi-18,SongzeLi-17,FanLi-19} focused on coded distributed computing (CDC) schemes to reduce the communication load (i.e., the number of information bits) of the Shuffle phase via coding, at the expense of increasing the computational load of the Map phase. The work \cite{Tao21} proposed a joint mapping and data shuffling scheme for a general heterogeneous CDC systems, with the aim of achieving a upper bound of the optimal communication load. Based on a low-rank optimization model for wireless MapReduce systems, the work \cite{Shi-19} aimed to maximize the achieved degree-of-freedom via building the interference alignment condition for data shuffling, where a difference-of-convex-function algorithm was developed. In the presence of full-duplex radios and imperfect channel state information, the work \cite{Simeone-19} proposed a superposition based scheme to simultaneously deliver coded multicasting massages and cooperatively precoded message in the Shuffle phase. Also, the work \cite{Feng21} investigated MapReduce over multihop device-to-device networks, and proposed an analog multi-level over-the-air (OTA) aggregation scheme for collecting IVAs in the Shuffle phase. The analog OTA aggregation technique that exploits the signal superposition property of wireless channels has been investigated in various applications such as IoT/sensor networks and edge machine learning\cite{Boche2013,Boche2015,Chen18,Zhu19,FengOTA20}. However, a homogeneous wireless network was assumed in all these works for MapReduce computation, and the communication cost in the Shuffle phase was assumed to be uniform.

 In reality, modern wireless communication networks usually consist of multiple radio access technologies (multi-RATs) as illustrated in Fig.~\ref{fig:SysMod}. For example, a 5G base station (i.e., gNB) and WiFi access points (APs) can be deployed in a cell, and a WD can use either the 5G network or WiFi network to access the internet. However, the associated communication costs across these multi-RAT networks are, in fact, different. For instance, the WiFi network (such as a campus network) may be free of charge but suffer from limited coverage. On the other hand, using the 5G RAT incurs a higher cost but benefits from a much broader coverage area. Most of the existing investigation works on the multi-RAT heterogeneous networks focused on improving the spectral/energy efficiency issues for generic wireless data transmissions\cite{YeLi-15,Olga-15,Hao17}. The work in \cite{XinWang-21} proposed an MEC-centric offloading decision scheme for the multi-RAT heterogeneous network, in which the WDs' usage cost and quality of service (QoS) are balanced by using the multi-armed bandit framework. Under the MapReduce framework, in addition to determining the dataset placement, transmission power, scheduling of the WDs and the computation tasks, we should also dynamically determine which RAT the WD should use to upload the IVAs for more efficiency over the multi-RAT networks. However, such aspects have not been explored before.


 In this paper, we study the dynamic resource allocation and RAT selection for MapReduce computation over multi-RAT networks. As shown in Fig.~\ref{fig:SysMod}, one 5G NB and several WiFi APs are deployed to provide wireless access services to multiple WDs in a single cell, and all the WDs are scheduled to collectively compute multiple data-processing functions under the MapReduce framework. The following summarizes the key contributions.
\begin{itemize}
    \item {\bf Dynamic Multi-RAT Selection for WDs:} The RAT selection provides an additional freedom for the WDs' IVA transmissions in achieving better system performance. The 5G gNB first performs dataset placement by splitting the dataset into multiple sub-datasets and assigning each to a WD. During the Shuffle phase of uploading the IVAs to the remote MEC server, each WD performs {\em RAT selection} by dynamically selecting either the 5G gNB or a WiFi AP to send IVAs to the MEC server over the air. The WiFi AP is low-cost in terms of data communication, but only a limited number of WDs can simultaneously access the WiFi AP due to its small coverage area. On the other hand, all the WDs in the cell can directly access the 5G gNB, but this incurs a higher data communication cost. Based on channel fading and cost variation, this RAT selection procedure is dynamic on a transmission time interval basis, and such dynamic selection enables a highly efficient MapReduce solution.

    \item{\bf OTA Aggregation for Nomographic Function Computing Under MapReduce Framework:} Under MapReduce framework for computing nomographic functions, the WDs are responsible for computing the Map functions based on their stored data files and producing IVAs, while the MEC server is responsible for computing the Reduce functions, whose input arguments are the {\em aggregated} IVAs. In order to enhance spectral efficiency for uploading IVAs to the MEC server, the analog OTA aggregation technique is employed to allow multiple WDs to simultaneously transmit their aggregated IVAs via the 5G and/or WiFi networks, by exploiting the signal superposition property of multiple access channels.

    \item{\bf Joint RAT Selection and Transceiver Optimization for Multi-RAT OTA Aggregation of IVAs:} We aim to minimize the weighted sum of the gNB/APs receivers' computational mean squared error (MSE) of the aggregated IVAs, the transmission cost, and the time delay in multi-RAT OTA aggregation of IVAs, which can provide a complete Pareto optimal solution set\cite{{Multi-Obj1}}. Under the energy budgets for the WDs' Map function computation and IVA transmission, we jointly optimize the RAT selection, the transmit coefficient per WD, and the gNB/APs' receive beamforming vectors for the reconstructed input of each Reduce function. Note that the RAT selection and the transceiver design for OTA aggregation are closely coupled in the MSE term and the transmit energy constraints. The formulated joint design problem is a nonconvex mixed-integer optimization problem.

    \item {\bf Efficient Algorithm:} Based on the minimum MSE principle, we obtain the optimal receiver beamforming vectors at the gNB/APs, and then transform the original problem into a joint RAT selection and transmit optimization problem. To handle the variable coupling and reduce computational complexity, we propose an efficient algorithm to obtain a stationary solution by employing Lagrange dual decomposition and primal decomposition to separate the original problem into two levels for optimizing the RAT selection and transmit coefficients, respectively. Extensive numerical results are provided to reveal the effectiveness of the proposed joint algorithm.
\end{itemize}



The remainder of the paper is organized as follows. Section II introduces the wireless MapReduce system model. Section III presents the RAT selection scheme for wireless MapReduce computing with OTA aggregation of IVAs. Section IV formulates the joint RAT selection and transceiver design problem to minimize the weighted sum of computational MSE, transmission cost, and time delay. Section V presents a low-complexity algorithm based on the Lagrange duality method and primal decomposition. Section VI provides numerical results to demonstrate the effectiveness of the proposed algorithm, followed by concluding remarks in Section VII. 

{\em Notation:} For an arbitrary-size matrix $\bm A$, $\bm A^T$ and $\bm A^H$ denote the transpose and Hermitian transpose, respectively. $\mathbb{R}^{x\times y}$ and $\mathbb{C}^{x\times y}$ denote the space of $x\times y$ matrices with complex and real entries, respectively. For a complex number $z$, $|z|$ denotes its absolute value, $z^\dagger$ denotes its conjugate, and ${\rm Re}[z]$ and ${\rm Im}[z]$ denote its real and imaginary parts, respectively. $\|\bm z\|$ denotes the Euclidean norm of a complex vector $\bm z$; $|{\cal X}|$ denotes the cardinality of a set $\cal X$. $\bm I$ and $\bm 0$ denote an identity matrix and an all-zeros vector/matrix, respectively, with appropriate dimensions; $x\sim {\cal CN}(\mu,\sigma^2)$ denotes the distribution of a circular symmetric complex Gaussian (CSCG) random variable $x$ with mean $\mu$ and variance $\sigma^2$, $x\sim {\cal U}[a,b]$ denotes the distribution of a uniform random variable $x$ within an interval $[a,b]$. Finally, $\mathbb{E}[\cdot]$ denotes the statistical expectation. 

\section{System Model}

\begin{figure}
\centering
  \includegraphics[width=3.5in]{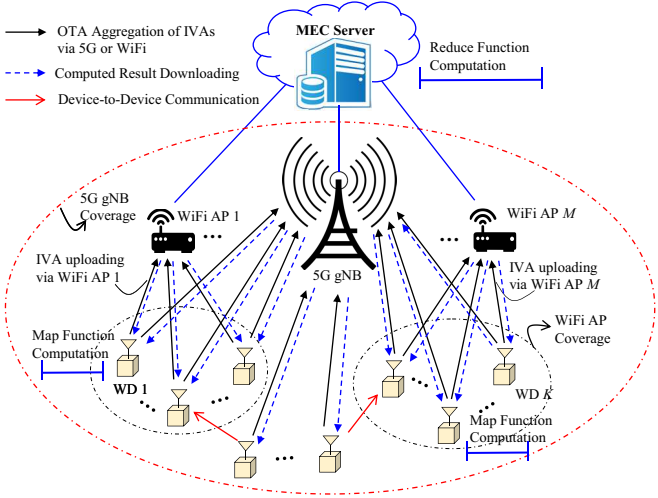}
  \caption{System model of wireless multi-RAT MapReduce computation.}\label{fig:SysMod}
\end{figure}

 We consider a wireless MapReduce computation system in which one 5G gNB and $M$ WiFi APs are deployed to serve $K$ WDs, as illustrated in Fig.~\ref{fig:SysMod}. The gNB and APs are connected to a common MEC server via optical fiber lines, where the data communication latency between the gNB/APs and the MEC server is negligible. The gNB and each AP are equipped with $N_{\rm 5g}$ and $N_{\rm wf}$ antennas, respectively, and each WD is equipped with a single antenna. The WDs are equipped with both 5G and WiFi interfaces so that they can communicate using either RAT. In addition, each WD is assumed to have communication, computing, and storage capabilities. Let ${\cal M} \triangleq \{1,...,M\}$ and ${\cal K}\triangleq \{1,...,K\}$ be the sets of APs and WDs, respectively.

 In this paper, we focus on the {\em nomographic function} computation job of processing a dataset which consists of $N$ data files $\{f_1,...,f_N\}$, where $N>K$. Denote by ${\cal N}\triangleq\{1,...,N\}$ the index set of the data files. Each data file $f_n\in\mathbb{F}_{2^{D}}$ is assumed to have a size of $D$ bits, $\forall n\in{\cal N}$. The $K$ WDs are scheduled to collectively compute a total of $Q$ nomographic functions $\{\phi_1,...,\phi_Q\}$, where each function $\phi_q:(\mathbb{F}_{2^D})^N \mapsto \mathbb{C}$ maps all the $N$ data files into a complex value $\phi_q(f_1,...,f_N)\in\mathbb{C}$. Typically, these nomographic functions $\{\phi_1,...,\phi_Q\}$ under consideration can be decomposed as\cite{Boche2013,Boche2015}
\begin{align}\label{eq.MapReduce-Decomp}
    \underbrace{\phi_q}_{\rm Nomographic~func.}(\underbrace{f_1,...,f_N}_{\rm Dataset}) &= \underbrace{h_q}_{\rm post-process.}\Big(\sum_{n=1}^N\underbrace{g_{q,n}}_{\rm pre-process.}(f_n)\Big) \notag \\
    &=h_q\left(\sum_{n=1}^N {\rm IVA}_{q,n}\right),~~\forall q\in{\cal Q},
\end{align}
 where $g_{q,n}:\mathbb{F}_{2^D}\mapsto \mathbb{C}$ denotes the pre-processing function which maps the input data file $f_n$ into a complex-valued IVA, i.e., ${\rm IVA}_{q,n}\triangleq g_{q,n}(f_n)\in\mathbb{C}$, $\forall n\in{\cal N}$, and $h_q: \mathbb{C}\mapsto\mathbb{C}$ denotes the post-processing function which maps the aggregated IVA associated with the $q$-th output function (i.e., $\sum_{n=1}^N{\rm IVA}_{q,n}$) into the value of nomographic function $\phi_q$.

 \begin{example}[Nomographic Function in WordCount Problem]
 In a typical WordCount problem in data analytics, one needs to count the number of occurrences of every word in a dataset including $N$ files $\{f_1,...,f_N\}$, where the different words are indexed by $q\in{\cal Q}$. The nomographic function $\phi_q$ is the number of occurrences of word $q$, which can be computed by executing the pre-processing function $g_{q,n}$ (to count the number of occurrences of word $q$ based on the file $f_n$) and post-processing function $h_q$ (to aggregate the pre-processed results based on the individual files).
 \end{example}

 \begin{example}[Nomographic Function in Sensor Networks]
 In environmental sensing/monitoring applications, some relevant statistics of sensor readings during a week or a month can be modelled as nomographic functions, such as the arithmetic mean ($\phi_q=\frac{1}{N}\sum_{n=1}^N \bm w^H f_n$ with $g_{q,n}(f_n)=\bm w^H f_n$ and $h_q(y) = y/N$), geometric mean ($\phi_q=(\prod_{n=1}^N \bm w^H f_n)^{1/N}$ with $g_{q,n}(f_n)=\log(\bm w^H f_n)$ and $h_q(y) = \frac{1}{N}\exp(y)$), and Euclidean norm ($\phi_q=\sqrt{\sum_{n=1}^N (\bm w^Hf_n)^2}$ with $g_{q,n}(f_n)=(\bm w^Hf_n)^2$ and $h_q(y) =\sqrt{y}$), where $\bm w$ serves as the data filtering vector.
 \end{example}

 \subsection{Distributed Wireless MapReduce Framework}
 Following the distributed MapReduce framework\cite{MapReduce-08}, the pre-processing functions $\{g_{q,n}\}_{q\in{\cal Q},n\in{\cal N}}$ and the post-processing functions $\{h_q\}_{q\in{\cal Q}}$ are referred to as the {\em Map} functions and the {\em Reduce} functions, respectively. As illustrated in Fig.~\ref{fig:SysMod}, the MEC server and the WDs are responsible for computation of the Reduce functions and Map functions, respectively. Due to the limited storage capacity for each WD, we suppose that the dataset is split and evenly distributed over the $K$ WDs. Denote by ${\cal N}_k\subseteq{\cal N}$ the index set of data files stored at WD $k$, $\forall k\in{\cal K}$. To fully utilize the storage resources of WDs, we assume that each data file $f_n$ is exclusively stored at one WD during the dataset placement phase. Since the data file number $N$ for the WDs' MapReduce computation is generally larger than the WD number $K$\cite{MapReduce-10,SongzeLi-17,SongzeLi-18}, it is assumed that $N/K$ can be treated as an integer number. Therefore, we have $n_k=|{\cal N}_k|=N/K$ data files per set ${\cal N}_k$, $\forall k\in{\cal K}$.

 Under the dataset placement strategy $\{{\cal N}_1,...,{\cal N}_K\}$, the MapReduce decomposition \eqref{eq.MapReduce-Decomp} of nomographic function $\phi_q$ can be re-expressed as
\begin{align}\label{eq.MapReduce-Decomp2}
    \phi_q(f_1,...,f_N) = h_q\left(\sum_{k=1}^K \sum_{n\in{\cal N}_k}{\rm IVA}_{q,n}\right),~~\forall q\in{\cal Q}.
\end{align}

 From \eqref{eq.MapReduce-Decomp2}, it follows that, under the MapReduce framework, each nomographic function $\phi_q$ can be collaboratively computed by the $K$ WDs via resorting to the computation of its associated Reduce function $h_q$, as well as the collection of the IVAs $\{{\rm IVA}_{q,n}\}_{n\in{\cal N}}$.  For each $n\in{\cal N}_k$, the ${\rm IVA}_{q,n}$ is generated as the output of WD $k$'s local computation for Map function $g_{q,n}(f_n)$. Instead of accessing all $N$ data files to directly compute the output function $\phi_q$, each ${\rm IVA}_{q,n}$ can be computed by one WD based on the data file $f_n$ in a distributed computing fashion. Clearly, the MapReduce decomposition \eqref{eq.MapReduce-Decomp2} includes two computation phases (Map function computation and Reduce function computation) and one communication phase (the collection of IVAs).

\begin{figure}
\centering
  \includegraphics[width=3.5in]{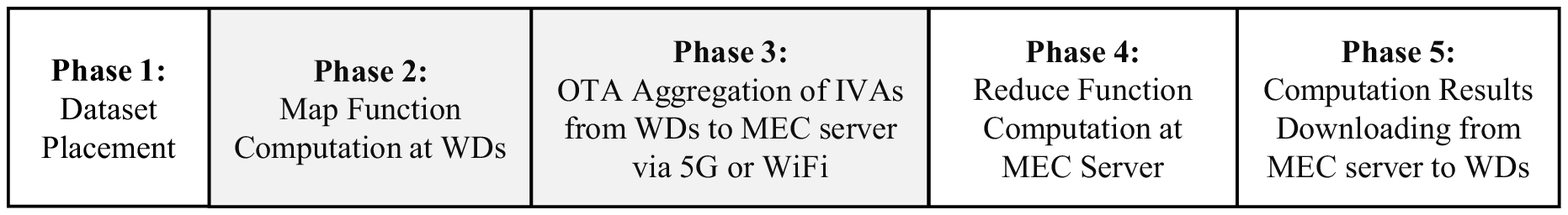}
  \caption{Proposed protocol for wireless multi-RAT MapReduce computation.}\label{fig:Protocol}
\end{figure}

\subsection{Proposed Wireless MapReduce Protocol}
 Fig.~\ref{fig:Protocol} presents the proposed wireless Mult-RAT MapReduce computation protocol for the $K$ WDs to compute the $Q$ nomographic functions $\{\phi_1,...,\phi_Q\}$ in a distributed fashion. It consists of five phases, introduced as follows.

\begin{itemize}
   \item{\em Dataset Placement Phase:} In this phase, the dataset $\{f_1,...,f_N\}$ is divided into $K$ disjoint subsets of equal cardinality, and each subset ${\cal N}_k$ of data files is assigned to be stored at each WD $k\in{\cal K}$. The index sets $\{{\cal N}_1,...,{\cal N}_K\}$ of the data files are determined for the $K$ WDs. 

   \item {\em WDs' Map Function Computation Phase:} In this phase, the $K$ WDs are enabled to locally compute the Map functions based on their respectively stored data files and generate the aggregated IVAs. Specifically, for each stored data file $f_n$ with $n\in{\cal N}_k$, WD $k$ completes the computation of the $Q$ Map functions $\{g_{1,n}(f_n),...,g_{Q,n}(f_n)\}$ and then outputs $Q$ IVAs, i.e., ${\rm IVA}_{1,n},...,{\rm IVA}_{Q,n}$, where $n\in{\cal N}_k$. By aggregating the obtained IVAs based on the nomographic function index $q\in{\cal Q}$, each WD $k\in{\cal K}$ obtains a set of aggregated IVAs $\{ \sum_{n\in{\cal N}_k}{\rm IVA}_{q,n},...,\sum_{n\in{\cal N}_k}{\rm IVA}_{Q,n}\}$.

   \item {\em WDs' IVA Uploading Phase:} In this phase, via accessing to either the 5G gNB or one WiFi AP, the $K$ WDs can upload their aggregated IVAs to the MEC server for the Reduce function computation therein. Since the 5G gNB and WiFi APs respectively operate in the licensed and unlicensed frequency bands, there exists no interference between the IVA transmission from the WDs to the 5G gNB and that to the WiFi AP.

   \item {\em MEC Server's Reduce Function Computation Phase:} In this phase, having obtained the aggregated IVAs from the WDs, the MEC server computes a total of $Q$ Reduce functions (i.e., $\{h_q\}_{q\in{\cal Q}}$) so as to generate the targeted nomographic function values $h_q(\sum_{k=1}^K\sum_{n\in{\cal N}_k}{\rm IVA}_{q,n}) = \phi_q(f_1,...,f_N)$, $q\in{\cal Q}$.

  \item {\em Computed Result Downloading Phase:} In this phase, each WD $k\in{\cal K}$ downloads the MEC server's computed results of the nomographic functions $\{\phi_q\}_{q\in{\cal Q}}$ by accessing to either the gNB or its associated WiFi AP.
\end{itemize}

 Motivated by the limited communication and computation resources of WDs, in this paper we focus on the energy consumption and communication cost of the $K$ WDs in the WDs' Map function computation phase and the communication phase of upload IVAs from the WDs to the MEC server.




\subsection{Map Function Computation at WDs}
 As illustrated in Fig.~\ref{fig:Protocol}, the $K$ WDs have the same time budget to complete the computation of their Map functions. We denote by $T_{\rm map}$ the time budget allocated for the $K$ WDs' Map function computation in parallel. Let $C_k$ denote the number of central processing unit (CPU) cycles required to compute one bit of a data file $f_n$ in WD $k$'s computation of the Map function $g_{q,n}(f_n)$, $\forall q\in{\cal Q}$, $n\in{\cal N}_k$. Recall that the size of each data file $f_n$ is $D$ bits. Therefore, each WD $k\in{\cal K}$ needs to execute a total of $QDC_kn_k$ CPU cycles to complete the computation of Map functions $\{g_{1,n},...,g_{Q,n}\}_{n\in{\cal N}_k}$, where $n_k=|{\cal N}_k|=N/K$. Without loss of generality, we assume that the data files are consecutively assigned to WD $k\in{\cal K}$, i.e., ${\cal N}_k=\{{(k-1)N/K+1},...,kN/K\}$. In order to successfully compute these $QN/K$ Map functions within the time duration $T_{\rm map}$, the CPU frequency of WD $k$ to execute each CPU cycle is adjusted as $f_k^{\rm cpu} = \frac{QDC_kN/K}{T_{\rm map}}$\cite{Feng18}. Accordingly, the amount of energy consumed by WD $k$'s Map function computation is given as \cite{Burd96,JunZhang17}
 \begin{align}
     E_k^{\rm map} &= \xi_kQDC_kn_k (f^{\rm cpu}_k)^2 = \frac{\xi_kQ^3D^3N^3C_k^3}{K^3T_{\rm map}^2},
 \end{align}
 where $\xi_k$ denotes the effective capacitance coefficient of WD $k$'s CPU chip architecture. At the end of the Map function computation phase, each WD $k\in{\cal K}$ has obtained the individual IVAs $\{{\rm IVA}_{q,n}\}_{q\in{\cal Q},n\in{\cal N}_k}$.

\section{RAT Selection for MapReduce with OTA Aggregation}

\begin{figure}
\centering
  \includegraphics[width=3.2in]{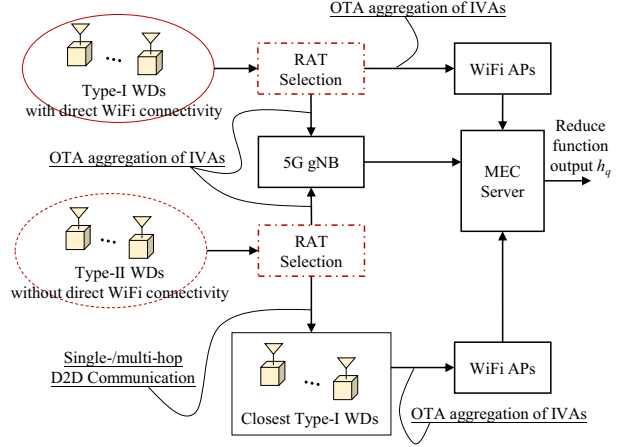}
  \caption{An illustration of RAT selection and OTA aggregation of IVAs via 5G or WiFi for wireless multi-RAT MapReduce computation systems.}\label{fig:RAT-sel}
\end{figure}

In this section, we first introduce the RAT selection of each WD, and then present the OTA aggregation of IVAs from the WDs to the MEC server via accessing the 5G or WiFi networks. Fig.~\ref{fig:RAT-sel} illustrates the $K$ WDs' RAT selection and OTA aggregation of IVAs via 5G or WiFi for wireless MapReduce computing.


\subsection{RAT Selection for IVA Transmission}
As discussed in the previous section, the individual IVAs $\{{\rm IVA}_{q,n}\}_{q\in{\cal Q},n\in{\cal N}_k}$ are locally generated at WD $k$ as the output of its Map functions $\{g_{1,n},...,g_{Q,n}\}_{n\in{\cal N}_k}$. Furthermore, by linearly combining the IVAs according to the Reduce function index $q\in{\cal Q}$, the aggregated IVAs of WD $k$, denoted by $x_{k,1},...,x_{k,Q}$, are locally generated at WD $k$, where $x_{k,q} \triangleq \sum_{n\in{\cal N}_k}{\rm IVA}_{q,n}$, $\forall k\in{\cal K}$, $q\in{\cal Q}$. Let a binary variable $\alpha_{k,q}\in\{0,1\}$ denote WD $k$'s RAT selection for uploading the aggregated IVA $x_{k,q}$ to the MEC server for performing the $q$-th Reduce function computation. For WD $k\in{\cal K}$ and time slot $q\in{\cal Q}$, we are ready to have
\begin{align}
    \alpha_{k,q} = \begin{cases}
    1, &{\rm if~WD}~k~{\rm uploads}~x_{k,q}~{\rm via~5G}\\
    0, &{\rm if~WD}~k~{\rm uploads}~x_{k,q}~{\rm via~WiFi}.
    \end{cases}
\end{align}

 \subsection{OTA Aggregation of IVAs from WDs to MEC Server via 5G}
 Denote by $\beta_{k,q}^{\rm 5g}\in\mathbb{C}$ the complex-valued transmit coefficient of WD $k\in{\cal K}$ during the transmission of aggregated IVA $x_{k,q}$ to the 5G gNB. Let $y_q^{\rm 5g}\in\mathbb{C}$ be the received signal at the 5G gNB. Note that all the $K$ WDs can directly communicate with the 5G gNB. Then, we have
 \begin{align}\label{eq.rate-5G}
     y_q^{\rm 5g} &= (\bm v^{\rm 5g}_{q})^H\left(\sum_{k\in{\cal K}}\alpha_{k,q} \bm h_{k,q}^{\rm 5g}\beta^{\rm 5g}_{k,q} x_{k,q} + \bm n_q\right),~\forall q\in{\cal Q},
 \end{align}
 where $\bm v_q\in\mathbb{C}^{N_{\rm 5g}\times 1}$ denotes the 5G gNB's receive aggregation vector; $\bm h_{k,q}^{\rm 5g}\in\mathbb{C}^{N_{\rm 5g}\times 1}$ denotes the channel coefficient vector for IVA transmission from WD $k$ to the 5G gNB; and $\bm n_{q}\sim {\cal CN}(0,\sigma^2\bm I_{N_{\rm 5g}})$ denotes the AWGN vector of the 5G gNB's receiver with zero-mean and variance $\sigma^2\bm I_{N_{\rm 5g}}$\cite{Goldsmith}.

 \subsection{OTA Aggregation of IVAs from WDs to MEC Server via WiFi APs}
 Let ${\cal K}^{\rm wf}_{m,q}=\{\pi_{m,q}(1),...,\pi_{m,q}(|{\cal K}^{\rm wf}_{m,q}|)\}\subseteq {\cal K}$ be the WD set associated with the $m$-th WiFi AP for implementing the $q$-th Reduce function, and its cardinality is denoted by $|{\cal K}^{\rm wf}_{m,q}|$. Due to the smaller coverage area of WiFi APs when compared to the 5G gNB, the whole cell cannot be fully covered by these $M$ APs, i.e., $\sum_{m=1}^M|{\cal K}^{\rm wf}_{m,q}|<K$, $\forall q\in{\cal Q}$. As a result, depending on whether they stay in the WiFi coverage, the $K$ WDs can be divided into the following two types:
 \begin{definition}[Type-I WD with Direct WiFi Connectivity] Each WD $k\in{\cal K}_q^{\rm wf}$ is referred to as a type-I WD under the WiFi coverage for the $q$-th Reduce function, if it can directly communicate with one WiFi AP, where ${\cal K}_q^{\rm wf}\triangleq \bigcup_{m=1}^M{\cal K}^{\rm wf}_{m,q}$.
 \end{definition}

 \begin{definition}[Type-II WD without Direct WiFi Connectivity]
 Each WD $k\in{\cal K}\setminus{\cal K}_q^{\rm wf}$ is referred to as a type-II WD for the $q$-th Reduce function, if it is out of the WiFi coverage and cannot directly communicate with one WiFi AP.
 \end{definition}

 We allow D2D communications for implementing the $q$-th Reduce function such that the type-II WD $j\in{\cal K}\setminus{\cal K}_q^{\rm wf}$ is allowed to communicate with another WD $k$ in its neighborhood. The communication network consisting of the $K$ WDs can be modelled as an undirected graph whose vertices are the WDs and whose edge set includes all the available D2D communication links among the $K$ WDs. Denote by ${\cal P}_{j\rightarrow k}$ the shortest path from WD $j\in{\cal K}$ to WD $k$. To guarantee the WiFi connectivity for each type-II WD $j\in{\cal K}\setminus{\cal K}_q^{\rm wf}$, we assume there exists at least one (single-hop or multi-hop) communication path to one type-I WD $k\in{\cal K}_q^{\rm wf}$. For a type-II WD $j\in{\cal K}\setminus{\cal K}_q^{\rm wf}$, the closest type-I WD is defined as $\delta_{j,q}$ such that
 \begin{align}
     \delta_{j,q} = \argmin_{k\in{\cal K}_q^{\rm wf}} |{\cal P}_{j\rightarrow k}|,~~\forall j\in {\cal K}\setminus{\cal K}^{\rm wf}_q,
 \end{align}
 where $|{\cal P}_{j\rightarrow k}|$ denotes the length (i.e., edge number) of path ${\cal P}_{j\rightarrow k}$ from WD $j$ to WD $k$.

 In the following, we introduce the OTA aggregation of IVAs via WiFi for type-I and type-II WDs, respectively. Fig.~\ref{fig:Two-types-WDs} illustrates the OTA aggregation of IVAs for type-I and type-II WDs.

\subsubsection{OTA Aggregation of IVAs for Type-I WDs via WiFi}
  As shown in Fig.~\ref{fig:Two-types-WDs}(a), for each type-I WD $k\in{\cal K}^{\rm wf}_{m,q}$, denote by $\beta^{\rm wf}_{k,q}\in\mathbb{C}$ its complex-valued transmit coefficient to send the aggregated IVA $x_{k,q}$ to its associated WiFi AP $m$ at the $q$-th time slot. Given the WD set ${\cal K}^{\rm wf}_{m,q}$ being associated with the $m$-th WiFi AP for the $q$-th Reduce function, the received signal $y_{m,q}^{\rm wf}\in\mathbb{C}$ of the $m$-th WiFi AP is expressed as
 \begin{align}\label{eq.rate-WiFi}
     y_{m,q}^{\rm wf} =& (\bm v^{\rm wf}_{m,q})^H\notag \\
     &~\times \Big(\sum_{k\in{\cal K}^{\rm wf}_{m,q}} (1-\alpha_{k,q})\bm h_{k,q}^{\rm wf}\beta^{\rm wf}_{k,q} x_{k,q} + \bm n_{m,q}\Big),
 \end{align}
 where $m=1,...,M$, $\bm v^{\rm wf}_{m,q}\in\mathbb{C}^{N_{\rm wf}\times 1}$ denotes the $m$-th WiFi AP's receive aggregation vector, $\bm h^{\rm wf}_{k,q}\in\mathbb{C}^{N_{\rm wf}\times 1}$ denotes the channel coefficient vector from WD $k\in{\cal K}_{m,q}^{\rm wf}$ to its associated WiFi AP $m$, and $\bm n_{m,q}\sim {\cal CN}(0,\sigma_m^2\bm I_{N_{\rm wf}})$ denotes the AWGN vector of WiFi AP $m$'s receiver.

\subsubsection{OTA Aggregation of IVAs for Type-II WDs via WiFi}
\begin{figure}
\centering
  \includegraphics[width=3.5in]{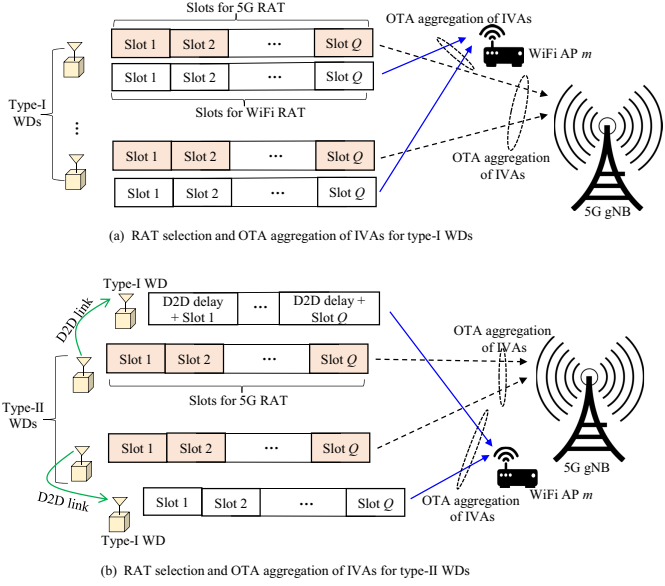}
  \caption{An illustration of RAT selection and OTA aggregation of IVAs for the two types of WDs.}\label{fig:Two-types-WDs}
\end{figure}

 As shown in Fig.~\ref{fig:Two-types-WDs}(b), for each type-II WD $j\in{\cal K}\setminus\bigcup_{m=1}^M{\cal K}_{m,q}^{\rm wf}$ and each Reduce function $h_q$, the number of time slots for IVA transmission to its closest type-I WD $\delta_{j,q}\in\bigcup_{m=1}^M{\cal K}_{m,q}^{\rm wf}$ is assumed to be equal to the number of edges $|{\cal P}_{j\rightarrow \delta_{j,q}}|$ of path ${\cal P}_{j\rightarrow \delta_{j,q}}$. Recall that the duration of each time slot is $T_{\rm tx}$ for one-hop D2D communication. As a result, there exists an additional time latency $T_{\rm tx}|{\cal P}_{j\rightarrow \delta_{j,q}}|$ for each type-II WD $j\in{\cal K}\setminus\bigcup_{m=1}^M{\cal K}_{m,q}^{\rm wf}$. For the $q$-th Reduce function, we denote by ${\cal K}^{\text{wf-d2d}}_{m,q}\subseteq{\cal K}_{m,q}^{\rm wf}$ the set of the closest neighbor type-I WDs associated with WiFi AP $m$ for type-II WDs' IVA uploading, and let the set ${\cal N}_{i,m,q}^{\rm II}\subseteq({\cal K}\setminus\cup_{m=1}^M{\cal K}_{m,q}^{\rm wf})$ collect all the type-II WDs which have a common closest neighbor type-I WD $i$ associated with WiFi AP $m$.

 Based on the OTA aggregation principle, the corresponding signal $y^{\text{wf-d2d}}_{m,q}$ received by WiFi AP $m$ is expressed as
 \begin{align}\label{eq.type2-y}
     & y^{\text{wf-d2d}}_{m,q}  = (\bm v_{m,q}^{\text{wf-d2d}})^H \notag \\
     & \times
      \Big( \sum_{i\in {\cal K}_{m,q}^{\text{wf-d2d}} }\bm h_{i,m}^{\text{wf-d2d}}\beta^{\text{wf-d2d}}_{i,q} (\sum_{j\in{\cal N}^{\rm II}_{i,m,q}}(1-\alpha_{j,q})x_{j,q}) + \bm n^{\text{wf-d2d}}_{m,q} \Big),
 \end{align}
 where $\beta_{i,q}^\text{wf-d2d}\in\mathbb{C}$ and $\bm h_{i,q}^\text{wf-d2d}\in\mathbb{C}^{N_{\rm wf}\times 1}$ denote the transmit coefficient and channel vector from WD $i\in{\cal K}_{m,q}^\text{wf-d2d}$ to its associated WiFi AP $m$, respectively; $\bm v_{m,q}^{\text{wf-d2d}}\in\mathbb{C}^{N_{\rm wf}\times 1}$ and $\bm n_{m,q}^{\text{wf-d2d}}\sim{\cal CN}(0,\sigma^2_m\bm I_{N_{\rm wf}})$ denote the receive aggregation vector and the AWGN vector at the WiFi AP $m$, respectively.

\subsection{Aggregated IVA Reconstruction at MEC Server}
 Let $\tilde{y}_q$ be the aggregated IVA to be reconstructed from the received signals $\{y_q,y^{\rm wf}_{m,q},y_{m,q}^{\text{wf-d2d}}\}$ at the gNB and APs, which serves as the input argument of the Reduce function $h_q$ for the MEC server. In order to approximate the targeted $\bar{x}_q = \sum_{n=1}^N x_{q,n}$, we employ the linear combination principle to aggregate the received signals $y_q^{\rm 5g}$ in \eqref{eq.rate-5G}, $y_{m,q}^{\rm wf}$ in \eqref{eq.rate-WiFi}, and $y_{m,q}^{\text{wf-d2d}}$ in \eqref{eq.type2-y}. As a result, the reconstructed $\tilde{y}_q$ is given as
\begin{align}\label{eq.tilde-y_q}
    \tilde{y}_q = y_q^{\rm 5g} + \sum_{m=1}^M y_{m,q}^{\rm wf} + \sum_{m=1}^M y_{m,q}^{\text{wf-d2d}},~~\forall q\in{\cal Q}.
 \end{align}
 Based on \eqref{eq.tilde-y_q}, the MEC server performs the Reduce function computation and generates the final results $\{ h_q(\tilde{y}_q) \}_{q\in{\cal Q}}$.

\section{Problem Formulation}
 In this section, we first elaborate the reward and penalty for 5G/WiFi RAT selection for the WDs' OTA aggregation of IVAs. Then we derive the computation MSE of the reconstructed input for the Reduce functions at the MEC server, and finally present the weighted MSE-cost-delay minimization problem.

\subsection{Reward and Penalty for 5G and WiFi RATs}

 \subsubsection{Energy Consumption for OTA Aggregation of IVAs}
 Let $E^{\rm tx}_{k,q}$ be the amount of transmit energy of WD $k$ for OTA aggregation of IVAs associated with the $q$-th Reduce function. Each type-I WD $k\in{\cal K}^{\text{wf-d2d}}_{m,q}$, in addition to uploading its own aggregated IVA to the MEC server via 5G or WiFi, is also responsible to upload the aggregated IVAs of the type-II WDs in set ${\cal N}_{m,k,q}^{\rm II}$. Therefore, based on the signal models \eqref{eq.rate-5G}, \eqref{eq.rate-WiFi}, and \eqref{eq.type2-y}, the amount of transmit energy of each WD $k\in{\cal K}_{m,q}^{\text{wf-d2d}}$ is given as
 \begin{align}\label{eq.E-typeI}
    E^{\rm tx}_{k,q} &=  ( \alpha_{k,q} |\beta^{\rm 5g}_{k,q}x_{k,q}|^2 + (1-\alpha_{k,q}) |\beta^{\rm wf}_{k,q}x_{k,q}|^2)T_{\rm tx} \notag \\
    &\quad \quad + \Big|\beta_{k,q}^{\text{wf-d2d}}  (\sum_{j\in{\cal N}^{\rm II}_{k,m,q}}(1-\alpha_{j,q})x_{j,q})\Big|^2T_{\rm tx}.
 \end{align}

 The type-I WD $k\in{\cal K}^{\rm wf}_{m,q}\setminus{\cal K}_{m,q}^{\text{wf-d2d}}$ is only responsible to send its aggregated IVA to the MEC server via 5G gNB or WiFi AP $m$, $\forall m\in{\cal M}$. Therefore, based on \eqref{eq.rate-5G} and \eqref{eq.rate-WiFi}, the amount of energy consumption for OTA aggregation of IVAs for WD $k\in{\cal K}_{m,q}^{\rm wf}\setminus{\cal K}_{m,q}^{\text{wf-d2d}}$ is given as \begin{align}\label{eq.E-typeI-minus}
     E^{\rm tx}_{k,q}= &(\alpha_{k,q} |\beta^{\rm 5g}_{k,q}x_{k,q}|^2 +  (1-\alpha_{k,q})|\beta^{\rm wf}_{k,q}x_{k,q}|^2)T_{\rm tx},\notag \\
     &\quad \quad\quad\quad\quad\quad\quad \forall k\in{\cal K}_{m,q}^{\rm wf}\setminus{\cal K}_{m,q}^{\text{wf-d2d}},m\in{\cal M}.
 \end{align}
 Each type-II WD $k\in{\cal K}\setminus(\cup_{m=1}^M{\cal K}^{\rm wf}_{m,q})$, meanwhile, can either select the 5G RAT for IVA uploading or transmit its aggregated IVA to its closest neighbour type-I WD $\delta_{k,q}\in\cup_{m=1}^M{\cal K}^{\rm wf}_{m,q}$ via D2D communication links with full transmit power $p^{\max}_k$. Therefore, for each type-II WD $k\in{\cal K}\setminus(\cup_{m=1}^M{\cal K}_{m,q}^{\rm wf})$, the amount of transmit energy for OTA aggregation of IVAs for the $q$-th Reduce function is given as
 \begin{align}\label{eq.E-typeII}
   E^{\rm tx}_{k,q} &= (\alpha_{k,q} |\beta^{\rm 5g}_{k,q}x_{k,q}|^2 + (1-\alpha_{k,q})p^{\max}_k) T_{\rm tx},\notag \\
   & \quad \quad\quad\quad\quad\quad\quad\quad\quad  \forall k\in{\cal K}\setminus(\cup_{m=1}^M{\cal K}_{m,q}^{\rm wf}).
 \end{align}

 \subsubsection{Cost for 5G and WiFi RATs}
 For each WD's OTA aggregation of IVAs, denote by $C_{\rm wf}$ and $C_{\rm 5g}$ the communication cost of accessing one WiFi AP and the 5G gNB, respectively, where the communication cost may be charged by the corresponding network service provider. Since WiFi operates over the unlicensed frequency bands, it is reasonably assumed that $C_{\rm wf}<C_{\rm 5g}$. Denote by $C_q$ the communication cost of the $K$ WDs to upload the aggregated IVAs which are associated with the $q$-th Reduce function. Based on the $K$ WDs' RAT selection profiles $\{\alpha_{k,q}\}_{k=1}^K$, their communication cost is expressed as
\begin{align}\label{eq.Cq}
    C_q = \sum_{k=1}^K\Big(\alpha_{k,q}C_{\rm 5g} + (1-\alpha_{k,q})C_{\rm wf}\Big),~~\forall q\in{\cal Q}.
\end{align}

\subsubsection{Time Delay Penalty for WiFi RAT}
 Note that the coverage of the 5G gNB is the whole cell. For each Reduce function $h_q$ to be computed by the MEC server, any WD $k\in{\cal K}$ can directly send its aggregated IVA $x_{k,q}$ to the MEC server within one time slot by accessing the 5G gNB. By employing the OTA aggregation method, the WDs selecting the 5G RAT can then simultaneously transmit their aggregated IVAs to the 5G gNB in one time slot. Therefore, the time delay for the OTA aggregation of IVAs associated with each Reduce function $h_q$ via 5G gNB is $t_q^{\rm 5g}=T_{\rm tx}$.

 On the other hand, the time delay for the OTA aggregation of IVAs via WiFi is dependent on the WD type. In the following, we consider type-I WDs with direct WiFi connectivity in set ${\cal K}_q^{\rm wf}$ and type-II WDs without direct WiFi connectivity ${\cal K}\setminus{\cal K}_q^{\rm wf}$ for the $q$-th Reduce function, respectively.
 \begin{itemize}
     \item For the type-I WDs to send their aggregated IVAs for one Reduce function computation at the MEC server, the time delay to implement the OTA aggregation of IVAs by accessing their associated WiFi APs is $t_q^{\rm wf}=T_{\rm tx}$.
    \item For the type-II WDs to select the WiFi RAT, due to the lack of direct WiFi connectivity, each type-II WD $j\in{\cal K}\setminus(\bigcup_{m\in{\cal M}}{\cal K}_{m,q}^{\rm wf})$ has to first transmit the aggregated IVA $x_{j,q}$ to its {\em closest} type-I WD $\delta_{j,q}\in{\cal K}_q^{\rm wf}$ via the shortest (single-hop or multi-hop) communication path ${\cal P}_{j\rightarrow \delta_{j,q}}$. The required communication time is proportional to the path length $|{\cal P}_{j\rightarrow \delta_{j,q}}|$ and is given as $|{\cal P}_{j\rightarrow \delta_{j,q}}|T_{\rm tx}$. After having received and decoded the aggregated IVAs from the type-II WDs, the closest type-I WDs employ the OTA aggregation of IVAs for the associated WiFi APs. As such, the total time delay penalty associated with the $q$-th Reduce function computation for IVA uploading from the type-II WDs to the MEC server via WiFi is expressed as
 \begin{subequations}\label{eq.t-wf-II}
 \begin{align}
    t^{\text{wf-d2d}}_q &= \max_{j\in{\cal K}\setminus{\cal K}_q^{\rm wf}} (1-\alpha_{j,q})|{\cal P}_{j\rightarrow{\delta_{j,q}}}|T_{\rm tx} + T_{\rm tx}\\
    &= (1-\alpha_{k_0^{(q)},q})|{\cal P}_{k_0^{(q)}\rightarrow{\delta_{k_0^{(q)},q}}}|T_{\rm tx} + T_{\rm tx},
 \end{align}
 \end{subequations}
  where the second term $T_{\rm tx}$ in (\ref{eq.t-wf-II}a) represents the time delay for the OTA aggregation from the closest type-I WDs to the WiFi APs (for the IVA uploading of type-II WDs), and (\ref{eq.t-wf-II}b) follows from $k_0^{(q)}\triangleq\argmax_{j\in{\cal K}\setminus{\cal K}_q^{\rm wf}} (1-\alpha_{j,q})|{\cal P}_{j\rightarrow{\delta_{j,q}}}|$.
 \end{itemize}

 It is important to note that the time delay for all the $K$ WDs' OTA aggregation of IVAs is dominated by the type-II WDs' IVA aggregation selecting the WiFi RAT, i.e., $t_q^{\text{wf-d2d}} > t_q^{\rm wf}=t_q^{\rm 5g}$. In this paper, we focus on the minimization of the time delay $t_q^{\text{wf-d2d}}$ for the $q$-th Reduce function, $\forall q\in{\cal Q}$.


\subsection{Computational MSE}
 Suppose that the Reduce function $h_q$ is {\em Lipschitz continuous} at point $\bar{x}_q = \sum_{k=1}^K x_{k,q}=\sum_{k=1}^K\sum_{n\in{\cal N}_k}{\rm IVA}_{q,n}$, $\forall q\in{\cal Q}$. This implies that there is a constant $C_0$ such that $|h_q(y)-h_q(x)|\leq C_0|y-x|$ for all $y\in\mathbb{C}$ sufficiently near $x$\cite{Scutari2017}. To measure the approximation performance of $h_q(\tilde{y}_q)$ with respect to the $q$-th Reduce function value $h_q(\bar{x}_q)$, we adopt the computational MSE between the obtained $\tilde{y}_q$ and the ground truth $\sum_{k=1}^Kx_{k,q}$ as the figure of merit for the $K$ WDs' OTA aggregation of IVAs, which is defined as
 \begin{subequations}\label{eq.mse-q}
 \begin{align}
     & {\rm MSE}_{q}\triangleq \mathbb{E}\left[ \Big|\tilde{y}_q - \sum_{k=1}^Kx_{k,q}\Big|^2 \right]\\
     &=\Big|(\bm v_q^{\rm 5g})^H \bm b_q^{\rm 5g} + \sum_{m=1}^M (\bm v_q^{\rm wf})^H\bm b_q^{\rm wf}+\sum_{m=1}^M(\bm v_q^\text{wf-d2d})^H\bm b_{m,q}^\text{wf-d2d}\notag \\ &-\sum_{k=1}^K x_{k,q}\Big|^2 + \|\bm v_{q}^{\rm 5g}\|^2\sigma^2  +  \sum_{m=1}^M \|\bm v_{m,q}^{\rm wf}\|^2\sigma_m^2 + \sum_{m=1}^M\|\bm v_{m,q}^{\text{wf-d2d}}\|^2\sigma_m^2\\
   &=\Big|\bm v_q^H \bm b_q - \sum_{k=1}^K x_{k,q}\Big|^2+\bm v_q^H\bm\Sigma_{n}\bm v_q,
  \end{align}
  \end{subequations}
  where $q=1,...,Q$, and the expectation $\mathbb{E}[\cdot]$ in (\ref{eq.mse-q}a) is taken over all the random AWGN terms. For \eqref{eq.mse-q}, we define
  \begin{align}
  &\bm v_q = [(\bm v_q^{\rm 5g})^T,(\bm v_{1,q}^{\rm wf})^T,...,(\bm v_{M,q}^{\rm wf})^T,(\bm v_{1,q}^{\rm wf-d2d})^T,...,(\bm v_{M,q}^{\rm wf-d2d})^T]^T \notag\\
    &\bm b_q = [(\bm b_q^{\rm 5g})^T,(\bm b_{1,q}^{\rm wf})^T,...,(\bm b_{M,q}^{\rm wf})^T,(\bm b_{1,q}^{\rm wf-d2d})^T,...,(\bm b_{M,q}^{\rm wf-d2d})^T]^T \notag\\
  &\bm \Sigma_n = {\rm diag}(\sigma^2\bm 1^T_{N_{\rm 5g}},\sigma_1^2\bm 1^T_{N_{\rm wf}},...,\sigma_M^2\bm 1^T_{N_{\rm wf}},\sigma_1^2\bm 1^T_{N_{\rm wf}},...,\sigma_M^2\bm 1^T_{N_{\rm wf}})\notag\\
  &\bm b^{\rm 5g}_q=\sum_{k=1}^K\alpha_{k,q}\beta_{k,q}^{\rm 5g}\bm h_{k,q}^{\rm 5g}x_{k,q} \label{eq.def-b-5g} \\
  &\bm b^{\rm wf}_{m,q} =\sum_{k\in{\cal K}_m^{\rm wf}}(1-\alpha_{k,q})\beta_{k,q}^{\rm wf}\bm h_{k,q}^{\rm wf}x_{k,q} \label{eq.def-b-wf}\\
  &\bm b^{\rm wf-d2d}_{m,q}=\sum_{k\in{\cal K}_m^\text{wf-d2d}}\beta_{k,q}^\text{wf-d2d}\bm h_{k,q}^\text{wf-d2d}(\sum_{j\in{\cal N}^{\rm II}_{m,k,q}}(1-\alpha_{j,q})x_{j,q}), \label{eq.def-b-wf-II}
\end{align}

 Note that under the fixed RAT decisions $\{\alpha_{k,q}\}$, the terms $\bm b^{\rm 5g}_{q}$, $\bm b^{\rm wf}_{m,q}$, and $\bm b^\text{wf-d2d}_{m,q}$ in \eqref{eq.def-b-5g}, \eqref{eq.def-b-wf}, and \eqref{eq.def-b-wf-II} are the linear functions of $\{\beta_{k,q}^{\rm 5g}\}$, $\{\beta^{\rm wf}_{k,q}\}$, and $\{\beta^\text{wf-d2d}_{k,q}\}$, respectively.

\subsection{MSE-Cost-Delay Minimization Problem}
 In this paper, subject to the energy constraints for the $K$ WDs' Map function computation and OTA aggregation of IVAs, our goal is to minimize the weighted sum of the computational MSE $\sum_{q=1}^Q{\rm MSE}_q$ in \eqref{eq.mse-q}, the communication cost $\sum_{q=1}^Q C_q$ in \eqref{eq.Cq}, and the time delay penalty $\sum_{q=1}^Q t_q^\text{wf-d2d}$ in \eqref{eq.t-wf-II}. We pursue the joint optimization of the RAT selection of the WDs $\{\alpha_{k,q}\}_{k\in{\cal K},q\in{\cal Q}}$ and the transceiver variables $(\{\beta^{\rm 5g}_{k,q}\}_{k\in{\cal K},q\in{\cal Q}},\{\bm v^{\rm 5g}_q\}_{q\in{\cal Q}})$ for all WDs' OTA aggregation of IVAs via 5G, and the transceiver variables $(\{\beta^{\rm wf}_{k_1,q}\}_{k_1\in(\bigcup_{m=1}^M{\cal K}_{q,m}^{\rm wf}),q\in{\cal Q}},\{\bm v^{\rm wf}_{m,q}\}_{m\in{\cal M},q\in{\cal Q}})$ for type-I WDs' OTA aggregation of IVAs via WiFi APs, the transceiver variables $(\{\beta_{k_2,q}^\text{wf-d2d}\}_{k_2\in(\bigcup_{m=1}^M{\cal K}_{m,q}^{\text{wf-d2d}}),q\in{\cal Q}},\{\bm v^{\text{wf-d2d}}_{m,q}\}_{m\in{\cal M},q\in{\cal Q}})$ for type-II WDs' OTA aggregation of IVAs via WiFi APs.

 Furthermore, we define the following three functions: $c_1(\{\alpha_{k,q},\bm \beta_q,\bm v_q\})\triangleq \sum_{q=1}^Q {\rm MSE}_q$, $c_2(\{\alpha_{k,q}\})\triangleq \sum_{q=1}^Q C_q$, and $c_3(\{\alpha_{k,q}\})\triangleq \sum_{q=1}^Q t_q^\text{wf-d2d}$, which correspond to the computational MSE, cost, and delay performance, respectively, under a certain RAT selection and transceiver design scheme. To achieve a tradeoff among the three objectives \cite{YeLi-15,Multi-Obj1,Multi-Obj2}, we formulate the following weighted sum minimization problem:
 \begin{subequations}\label{eq.prob1}
\begin{align}
   &({\rm P}1): ~~ \min_{\{\bm \alpha_{q},\bm \beta_q,\bm v_q\}}~\sum_{j=1}^3\omega_j c_j
   \\
    &~~{\rm s.t.}~~\alpha_{k,q}\in\{0,1\},~\forall k\in{\cal K},~ q\in{\cal Q}\\
    &~~\quad\quad E_k^{\rm map} + \sum_{q=1}^Q E_{k,q}^{\rm tx} \leq E_k,~ \forall k\in{\cal K},
\end{align}
\end{subequations}
 where $\bm \alpha_q\triangleq[\alpha_{1,q},...,\alpha_{K,q}]^T$ collects all the WDs' binary RAT selection variables; ${\bm \beta}_q\triangleq\{\beta_{k,q}^{\rm 5g},\beta_{k_1,q}^{\rm wf},\beta^{\text{wf-d2d}}_{k_2,q}\}$ collects all the WDs' transmit coefficients in OTA aggregation of IVAs with $k\in{\cal K}$, $k_1\in{\cal K}_{m,q}^{\rm wf}$, $k_2\in{\cal K}_{m,q}^{\text{wf-d2d}}$, $m\in{\cal M}$, and $q\in{\cal Q}$; and $\omega_j\geq 0$ denotes the nonnegative weight for $j\in\{1,2,3\}$, which specifies the priority of the $j$-th objective and reflects the system operator's preference. By varying the weights $\{\omega_j\}_{j=1}^3$, we can obtain a complete Pareto optimal set which corresponds to a set of RAT selection and transceiver designs for OTA aggregation of IVAs. In this paper, we refer to problem (P1) as the weighted sum of MSE-cost-delay (WS-MCD) minimization problem.

 For the WS-MCD problem (P1), the constraints in (\ref{eq.prob1}b) denote the binary RAT selections in uploading their aggregated IVAs to the MEC server, and the $k$-th constraint of (\ref{eq.prob1}c) represents that the amount of energy consumed due to WD $k$'s Map function computation and IVA transmission cannot exceed its energy budget $E_k$. Note that, due to the variable coupling in the $\{{\rm MSE}_{q}\}_{q\in{\cal Q}}$ and the binary variables $\{\alpha_{k,q}\}_{k\in{\cal K},q\in{\cal Q}}$, problem (P1) is a mixed-integer and non-convex optimization problem.

\section{Proposed Solution for Problem (P1)}
 In this section, we present the proposed joint RAT selection and transceiver design solution for problem~(P1).

\subsection{Problem Transformation for Problem (P1)}

 Note that the constraints (\ref{eq.prob1}c) and (\ref{eq.prob1}d), as well as the cost functions $\{C_q\}$ and the delay functions $\{t_q^\text{wf-d2d}\}$ in the objective function of problem (P1), are independent of the receive aggregation vectors $\{\bm v_{q}\}$. In addition, the ${\rm MSE}_q$ given in (\ref{eq.mse-q}c) is a convex quadratic function of $\bm v_q$, and there exists no coupling between ${\rm MSE}_{q_1}$ and ${\rm MSE}_{q_2}$ for $q_1\neq q_2\in{\cal Q}$. Therefore, we can obtain the optimal solution of $\{\bm v_q\}$ for problem (P1) by setting the first-order derivative of ${\rm MSE}_q$ with respect to $\bm v_q$ to be zero. Formally, we establish the following lemma on the optimal receive aggregation vectors at gNB/APs to minimize the sum MSE $c_1=\sum_{q=1}^Q{\rm MSE}_q$.

 \begin{lemma}[Optimal Receive Aggregation Vectors for MSE minimization] \label{lem.BF}
Under the given $\{\alpha_{q},\bm \beta_{q}\}$, the optimal solution $\bm v^*_q = \{\bm v_{q}^{\rm 5g*}, \bm v_{k,q}^{\rm wf*}, \bm v_{m,q}^\text{wf-d2d*}\}$ to minimize $c_1=\sum_{q=1}^Q{\rm MSE}_q$ is expressed as
\begin{subequations}
\begin{align*}
 &   \bm v_q^{\rm 5g*} = [{\bm v}^*_q]_{1:N_{\rm 5g}}\\
 &   \bm v_{m,q}^{\rm wf*} = [{\bm v}^*_q]_{(N_{\rm 5g}+(m-1)N_{\rm wf}):(N_{\rm 5g}+mN_{\rm wf})} \\
 &   \bm v_{m,q}^\text{wf-d2d*} = [{\bm v}^*_q]_{(N_{\rm 5g}+(M+m-1)N_{\rm wf}):(N_{\rm 5g}+(M+m)N_{\rm wf})},
\end{align*}
\end{subequations}
where $m\in{\cal M}$, $q\in{\cal Q}$, and
\begin{subequations}\label{eq.Opt-BF}
\begin{align}
    {\bm v}^*_q &= \argmin_{{\bm v}_{q}}~ {\rm MSE}_q\\
    &=\Big(\sum_{k=1}^K x_{k,q}\Big)^{\dagger}({\bm b}_q {\bm b}_q^H + \bm \Sigma_n)^{-1}{\bm b}_q,~~\forall q\in{\cal Q}.
\end{align}
\end{subequations}
\end{lemma}

\begin{IEEEproof}
 Due to the Hessian matrix ${\bm b}_q{\bm b}_q^H+\bm \Sigma_n\succ \bm 0$ of the ${\rm MSE}_q$ with respect to ${\bm v}_q$ being positive definite, the function ${\rm MSE}_q$ is a strictly convex function of variables $\{{\bm v}_q\}$. Therefore, the optimal solution of $\{{\bm v}^*_q\}$ for minimizing ${\rm MSE}_q$ is given by the linear minimum MSE (LMMSE) or the Wiener filter\cite{Kay93}. Specifically, by setting the first-order derivative of ${\rm MSE}_q$ with respect to ${\bm v}_q$ to be zero (i.e., $\nabla_{\bm v_q}{\rm MSE}_q=\bm 0$), the optimal ${\bm v}^*_q$ is obtained as a function of the variable vector $\bm b_q$; i.e., ${\bm v}^*_q = \Big(\sum_{k=1}^K x_{k,q}\Big)^{\dagger}({\bm b}_q{\bm b}_q^H + \bm \Sigma_n)^{-1}{\bm b}_q$, $\forall q\in{\cal Q}$.

 Furthermore, based on the definition of $\bm v_q$, the optimal $\{\bm v_{q}^{\rm 5g*}, \bm v_{m,q}^{\rm wf*}, \bm v_{m,q}^\text{wf-d2d*}\}$ is obtained as shown in Lemma~\ref{lem.BF}.
\end{IEEEproof}

 By substituting $\{\bm v^*_{q}\}$ into (\ref{eq.mse-q}b), the ${\rm MSE}_q$ is re-expressed as a function of $(\bm \alpha_q,\bm \beta_q)$
\begin{subequations}\label{eq.MSE-v-opt}
 \begin{align}
     &{\rm MSE}_q(\bm \alpha_q, \bm \beta_q) =\Big|\sum_{k=1}^Kx_{k,q}\Big|^2(1-\bm b_q^H(\bm b_q\bm b_q^H+\bm \Sigma_n)^{-1}\bm b_q)\notag \\
     &\quad\quad=\frac{\big|\sum_{k=1}^K x_{k,q}\big|^2}{1 + \bm b_q^H\bm \Sigma_n^{-1} \bm b_q} \\
     &\quad\quad= \frac{\big|\sum_{k=1}^K x_{k,q}\big|^2}{1+
     \frac{\|\bm b^{\rm 5g}_q\|^2}{\sigma^2} + \sum_{m=1}^M \frac{\|\bm b^{\rm wf}_{m,q}\|^2}{\sigma^2_m} + \sum_{m=1}^M \frac{\|\bm b^\text{wf-d2d}_{m,q}\|^2}{\sigma^2_m}},
 \end{align}
 \end{subequations}
 where (\ref{eq.MSE-v-opt}a) holds from the Sherman-Morrison formula \cite{Matrix-Inverse} $(\bm A+ \bm u\bm v^H)=\bm A^{-1}-\frac{\bm A^{-1}\bm u\bm v^H\bm A^{-1}}{1+\bm v^H\bm A^{-1}\bm u}$ by setting $\bm A=\bm \Sigma_n$ and $\bm u=\bm v = \bm b_q$, and (\ref{eq.MSE-v-opt}b) follows from the definitions $\bm b^{\rm 5g}_q$, $\bm b^{\rm wf}_q$, and $\bm b^\text{wf-d2d}_q$ in \eqref{eq.def-b-5g}, \eqref{eq.def-b-wf}, and \eqref{eq.def-b-wf-II}, respectively.

\begin{lemma}[Optimal Active Energy Constraints for (P1)]\label{lem-active}
 At the optimality of problem (P1), each WD's energy constraint in (\ref{eq.prob1}d) becomes active, i.e,
\begin{align}
    E_k^{\rm map} + \sum_{q=1}^Q E_{k,q}^{\rm tx} = E_k,~\forall k\in{\cal K}.
\end{align}
\end{lemma}
\begin{IEEEproof}
Based on \eqref{eq.MSE-v-opt}, it can be verified that the ${\rm MSE}_q$ monotonically decreases with $\|\bm b^{\rm 5g}_q\|^2$, $\|\bm b^{\rm wf}_q\|^2$, and $\|\bm b^\text{wf-d2d}_q\|^2$. From \eqref{eq.def-b-5g}, it follows that the value $\|\bm b^{\rm 5g}_q\|=\|\sum_{k=1}^K\alpha_{k,q}\beta_{k,q}^{\rm 5g}\bm h_{k,q}^{\rm 5g}x_{k,q}\|$ monotonically increases with each $|\beta_{k,q}^{\rm 5g}|^2$, $\forall k$. Likewise, from \eqref{eq.def-b-wf} and \eqref{eq.def-b-wf-II}, it is shown that the values $\|\bm b^{\rm wf}_{m,q}\| = \|\sum_{k\in{\cal K}_m^{\rm wf}}(1-\alpha_{k,q})\beta_{k,q}^{\rm wf}\bm h_{k,q}^{\rm wf}x_{k,q}\|$ and $\|\bm b^{\rm wf-d2d}_{m,q}\|=\|\sum_{k\in{\cal K}_m^\text{wf-d2d}}\beta_{k,q}^\text{wf-d2d}\bm h_{k,q}^\text{wf-d2d}\sum_{j\in{\cal N}^{\rm II}_{m,k,q}}(1-\alpha_{j,q})x_{j,q}\|$ both increase with the increasing of each $|\beta_{k,q}^{\rm wf}|^2$ and $|\beta_{k,q}^\text{wf-d2d}|^2$, $\forall k$, respectively. Note that the term $|\beta_{k,q}^{\rm 5g}|^2$ corresponds to the transmit energy of WD $k\in{\cal K}$ in sending its aggregated IVA to the 5G gNB, $|\beta_{k,q}^{\rm wf}|^2$ corresponds to the transmit energy of WD $k\in{\cal K}_m^{\rm wf}$ to the WiFi AP $m$, and $|\beta_{k,q}^\text{wf-d2d}|^2$ corresponds to the transmit energy of WD $k\in{\cal K}_m^\text{wf-d2d}$ to the WiFi AP $m$. This implies that one can always increase the transmit energy of WDs to decrease the ${\rm MSE}_q$, thereby leading to a smaller value of the objective function of problem (P1). Therefore, by contradiction, it is yielded that each constraint of (\ref{eq.prob1}d) must hold with {\em strict} equality in order to minimize the weighted sum of the MSE, cost, and delay for problem (P1).
\end{IEEEproof}

For the ${\rm MSE}_q(\bm \alpha_q,\bm \beta_q)$ in \eqref{eq.MSE-v-opt}, we establish its non-convexity as follows.
\begin{lemma}[Non-convexity of ${\rm MSE}_q(\bm \alpha_q,\bm \beta_q)$]\label{lem.nonconvex}
The expression of the ${\rm MSE}_q(\bm \alpha_q,\bm \beta_q)$ in \eqref{eq.MSE-v-opt} is non-convex in each component of $\bm \alpha_{q}$ and $\bm \beta_q$.
\end{lemma}

\begin{IEEEproof}
Lemma~\ref{lem.nonconvex} can be verified by checking the non-convexity of the function $f(x)=\frac{b_0}{1+c_0+x^2}$ in $x\geq0$, where $b_0\geq 0$ and $c_0\geq 0$ are constant terms, and the second-order derivative of $f(x)$ is given as $f^{\prime\prime}(x)=\frac{2b_0(3x^2-1-c_0)}{(1+c_0+x^2)^3}$. It is not always guaranteed that $f^{\prime\prime}(x)\geq 0$ for all $x\geq 0$. Based on the non-negativeness property of the second-order derivative for a convex function, the function $f(x)=\frac{b_0}{1+c_0+x^2}$ is non-convex in $x\geq 0$. Since the left-hand-side of (\ref{eq.MSE-v-opt}b) can be written as the form of $\frac{b_0}{1+c_0+x^2}$, it is thus shown that the ${\rm MSE}_q(\bm \alpha_q,\bm \beta_q)$ is a non-convex function in each component of $\bm \alpha_{q}$ or $\bm \beta_q$.
\end{IEEEproof}

Furthermore, the variables $\bm \alpha_q$ and $\bm \beta_q$ are coupled in ${\rm MSE}_q(\bm \alpha_q,\bm \beta_q)$. Based on \eqref{eq.MSE-v-opt} and Lemma~\ref{lem-active}, the original problem (P1) is reduced into
\begin{subequations}\label{eq.prob2}
\begin{align}
{\rm (P2):}  &\min_{\{\bm \alpha_{q},\bm \beta_q\}}~~\sum_{q=1}^Q \left(\omega_1 \frac{|\sum_{k=1}^Kx_{k,q}|^2}{1 + \bm b_q^H\bm \Sigma^{-1}_n \bm b_q} + \omega_2 C_q + \omega_3 t_q^\text{wf-d2d}\right)
 \\
    &{\rm s.t.}~~\alpha_{k,q}\in\{0,1\},\forall k\in{\cal K},q\in{\cal Q}\\
    & \quad\quad \sum_{q=1}^Q p_{k,q}^{\rm tx}(\bm \alpha_q,\bm \beta_q) = \tilde{p}_k,~\forall k\in{\cal K},
\end{align}
\end{subequations}
where $\tilde{p}_k\triangleq \frac{E_k}{T_k}-\frac{\xi_kQ^3D^3N^3C_k^3}{K^3T_{\rm map}^2T_{\rm tx}}$ and $p^{\rm tx}_{k,q}(\bm \alpha_q,\bm \beta_q)$ is defined as
\begin{align}
  p^{\rm tx}_{k,q}(\bm \alpha_q,\bm \beta_q) \triangleq  &  \begin{cases}
    &\alpha_{k,q}|\beta_{k,q}^{\rm 5g}x_{k,q}|^2+(1-\alpha_{k,q})p_k^{\max},\\
    & \quad\quad\quad\quad\quad {\rm if}~k\in {\cal K}\setminus(\cup_{m=1}^M{\cal K}_{m,q}^\text{wf})\\
    &\alpha_{k,q}|\beta_{k,q}^{\rm 5g}x_{k,q}|^2 + (1-\alpha_{k,q})|\beta_{k,q}^{\rm wf}x_{k,q}|^2 , \\
    &\quad\quad\quad\quad\quad {\rm if}~ k\in {\cal K}_{m,q}^{\rm wf}\setminus{\cal K}_{m,q}^\text{wf-d2d},m\in{\cal M}\\
    &\alpha_{k,q}|\beta_{k,q}^{\rm 5g}x_{k,q}|^2 + (1-\alpha_{k,q})|\beta_{k,q}^\text{wf}x_{k,q}|^2  \\
    & ~ + |\beta_{k,q}^\text{wf-d2d}(\sum_{j\in{\cal N}^{\rm II}_{k,m,q}}(1-\alpha_{j,q})x_{j,q})|^2,\\
    &\quad\quad\quad\quad\quad {\rm if}~ k\in{\cal K}_{m,q}^\text{wf-d2d},m\in{\cal M}.
\end{cases}
\end{align}
For problem (P2), besides the RAT association vector $\{\bm \alpha_{q}\}$, the coupling of variable vectors $\{\bm \alpha_{q}\}$ and $\{\bm \beta_q\}$ in (\ref{eq.prob2}b--c) makes (P2) highly challenging to solve. We next employ the Lagrange duality method to find a stationary point for (P2).

\subsection{Proposed Solution for (P2)}
For problem (P2), the associated {\em partial} Lagrangian is expressed as
\begin{subequations}\label{eq.Lagrange_func}
\begin{align}
  {\cal L}(\{\bm \alpha_q,\bm \beta_q\},\bm \lambda) =& \sum_{q=1}^Q \left(\omega_1 \frac{|\sum_{k=1}^Kx_{k,q}|^2}{1 + \bm b_q^H\bm \Sigma^{-1}_n \bm b_q} + \omega_2 C_q + \omega_3 t_q^\text{wf-d2d}\right)  \notag \\
 &  + \sum_{k=1}^K\lambda_k(\sum_{q=1}^Qp_{k,q}^{\rm tx}(\bm \alpha_q,\bm \beta_q) - \tilde{p}_k)\\
=& \sum_{q=1}^Q g_q(\bm \alpha_q,\bm \beta_q,\bm \lambda) - \sum_{k=1}^K \lambda_k\tilde{p}_k,
\end{align}
\end{subequations}
where $\bm \lambda\triangleq[\lambda_1,...,\lambda_K]^T$, with $\lambda_k$ denoting the Lagrange multiplier for the $k$-th constraints in (\ref{eq.prob2}c), and the function $g_q(\bm \alpha_q,\bm \beta_q,\bm \lambda)$ in (\ref{eq.Lagrange_func}b) is defined as
\begin{align}
 g_q(\bm \alpha_q,\bm \beta_q,\bm \lambda)
 \triangleq &  \sum_{k=1}^K\lambda_kp_{k,q}^{\rm tx}(\bm \alpha_q,\bm \beta_q) \notag\\
 & + \frac{\omega_1|\sum_{k=1}^Kx_k|^2}{1+\bm b_q^H\bm \Sigma_n^{-1}\bm b_q}  + \omega_2C_q + \omega_3t_q^\text{wf-d2d},
\end{align}
where $q\in{\cal Q}$. Accordingly, the Lagrange dual function of problem (P2) is defined as
\begin{align}\label{eq.dual-func}
    {\cal D}(\bm \lambda) \triangleq & \min_{\{\bm \alpha_q, \bm \beta_q\}}{\cal L}(\{\bm \alpha_q, \bm \beta_q\},\bm \lambda)\notag \\
    &{\rm s.t.}~\alpha_{k,q}\in\{0,1\},~\forall k\in{\cal K},q\in{\cal Q}.
\end{align}
The dual problem of (P2) is given as
\begin{align}\label{eq.prob-dual}
    &\max_{\bm \lambda}~{\cal D}(\bm \lambda).
\end{align}

Under the given dual variable vector $\bm \lambda$, the Lagrangian ${\cal L}(\{\bm \alpha_q,\bm \beta_q\},\bm \lambda)$ in \eqref{eq.Lagrange_func} can be further decomposed $Q$ subproblems; i.e., the joint RAT selection and transmit optimization problems $({\rm P3}.q)$ with $q\in{\cal Q}$ as follows:
\begin{subequations}
\begin{align}
    ({\rm P3}.q): \min_{\bm \alpha_q,\bm \beta_q}&~{g_q(\bm \alpha_q,\bm \beta_q,\bm \lambda)} \\
    {\rm s.t.}&~\alpha_{k,q}\in\{0,1\},~\forall k\in{\cal K}.
\end{align}
\end{subequations}

\subsection{Joint RAT Selection and Transmit Optimization}
We first focus on finding the solution of the joint RAT selection and transmit optimization problem (${\rm P}3.q$) for all $q\in{\cal Q}$. Since the RAT selection vector $\bm \alpha_q$ and the transmit coefficient vector $\bm \beta_q$ are coupled, the primal decomposition method \cite{Boyd} is employed to separate problem (P3.$q$) into the following two levels of optimization problem. By fixing the RAT selection vector $\bm \alpha_q$ in problem (P3.$q$), we obtain the transmit coefficient optimization problem as
\begin{align}\label{eq.prob-Tx}
\min_{\bm \beta_q}~
\sum_{k=1}^K\lambda_kp_{k,q}^{\rm tx}(\bm \alpha_q,\bm \beta_q) + \frac{\omega_1|\sum_{k=1}^Kx_k|^2}{1+\bm b_q^H\bm \Sigma_n^{-1}\bm b_q},
\end{align}
where the two terms $\omega_2C_q$ and $\omega_3t_q^\text{wf-d2d}$ in $g_q(\bm \alpha_q,\bm \beta_q,\bm \lambda)$ are irrelevant to $\bm \beta_q$ and thus removed in problem \eqref{eq.prob-Tx}. On the other hand, by fixing the transmit coefficient vector $\bm \beta_q$ in problem (P3.$q$), we obtain the RAT selection problem as
\begin{align}\label{eq.prob-RAT}
 &\min_{\bm \alpha_q} \sum_{k=1}^K\lambda_kp_{k,q}^{\rm tx}(\bm \alpha_q,\bm \beta_q) + \frac{\omega_1|\sum_{k=1}^Kx_k|^2}{1+\bm b_q^H\bm \Sigma_n^{-1}\bm b_q} + \omega_2C_q + \omega_3t_q^\text{wf-d2d}\notag\\
 &~~{\rm s.t.}~~\alpha_{k,q}\in\{0,1\},~\forall k\in{\cal K}.
\end{align}

In the following, we respectively present the solutions of problems \eqref{eq.prob-Tx} and \eqref{eq.prob-RAT}.
\subsubsection{Solving Transmit Optimization Problem \eqref{eq.prob-Tx}}
As in previous discussion, the second term in the objective function of problem \eqref{eq.prob-Tx} is non-convex with respect to $\bm \beta_q$. Then, we pursue obtaining a local optimal solution for \eqref{eq.prob-Tx} by letting the first-order derivative of the objective function be zero.
\begin{lemma}[Transmit Coefficient Solution $\bm \beta_q^*$]\label{lem.opt-beta}
The optimal $\bm \beta_q^*$ for problem \eqref{eq.prob-Tx}  satisfies the following equations:
\begin{subequations}\label{eq.sol-Tx}
\begin{align}
    & \frac{\omega_1|\sum_{k=1}^Kx_{k,q}|^2}{(1+\bm b_q^H\bm \Sigma_n^{-1}\bm b_q)^2}\frac{(\bm b_q^{\rm 5g})^H\bm h^{\rm 5g}_{k,q}\alpha_{k,q}x_{k,q}}{\sigma^2}  \notag \\
    &\quad\quad\quad - \lambda_k\alpha_{k,q}\beta^{\rm 5g}_{k,q}|x_{k,q}|^2 = 0,~\forall k\in{\cal K}\\
    & \frac{\omega_1|\sum_{k=1}^K x_{k,q}|^2}{(1+\bm b_q^H\bm \Sigma_n^{-1}\bm b_q)^2}\frac{(\bm b_q^{\rm wf})^H\bm h^{\rm wf}_{k,q}(1-\alpha_{k,q})x_{k,q}}{\sigma_m^2} \notag \\
    &\quad \quad\quad - \lambda_{k}(1-\alpha_{k,q})\beta_{k,q}^{\rm wf}|x_{k,q}|^2 = 0,~\forall k\in{\cal K}_{m,q}^{\rm wf}\\
    & \frac{\omega_1|\sum_{k=1}^K x_{k,q}|^2}{(1+\bm b_q^H\bm \Sigma_n^{-1}\bm b_q)^2} \frac{(\bm b_q^\text{wf-d2d})^H\bm h^\text{wf-d2d}_{k,q} \Big(\sum_{j\in{\cal N}_{m,k,q}^{\rm II}}(1-\alpha_{j,q})x_{j,q}\Big)}{\sigma_m^2} \notag \\
    &- \lambda_{k}\beta_{k,q}^\text{wf-d2d} \big|\sum_{j\in{\cal N}_{m,k,q}^{\rm II}}(1-\alpha_{j,q})x_{j,q}\big|^2 = 0,~\forall k\in {\cal K}_{m,q}^\text{wf-d2d}.
\end{align}
\end{subequations}
\end{lemma}

\subsubsection{Solving RAT Selection Problem \eqref{eq.prob-RAT}}
Due to the integer constraints and the nonconvexity of $\frac{\omega_1|\sum_{k=1}^Kx_k|^2}{1+\bm b^H_q \bm \Sigma_n^{-1}\bm b_q}$ in $\bm \alpha_q$, problem \eqref{eq.prob-RAT} is a non-convex mixed-integer problem. To tackle this difficulty, we relax the integer variables $\{\alpha_{k,q}\}$ into continuous ones $\{\tilde{\alpha}_{k,q}\}$, such that $0\leq \tilde{\alpha}_{k,q}\leq 1$, $\forall k\in{\cal K}$. Similarly, based on the first-order derivative conditions, the relaxed WDs' RAT selection variables can be obtained.
\begin{lemma}[Relaxed RAT Selection Solution $\{\tilde{\alpha}^*_{k,q}\}$]\label{lem.Relaxed-alpha}
The relaxed RAT selection variables $\{\tilde{\alpha}^*_{k,q}\}$ can be obtained as $\tilde{\alpha}^*_{k,q}=[\hat{\alpha}^*_{k,q}]_0^{1}$, $\forall k$, where $[x]_0^1$ denotes the convex projection mapping $x$ into the interval $[0,1]$ such that $[x]_0^1=x$ for $x\in[0,1]$, $[x]_0^1=1$ for $x>1$, and $[x]_0^1=0$ for $x<0$. The optimal $\{\hat{\alpha}^*_{k,q}\}$ satisfy the following equations:
\begin{subequations}\label{eq.Rel-alpha}
\begin{align}
&  \frac{2 w_1|\sum_{k=1}^Kx_{k,q}|^2}{(1+\bm b_q^H\bm \Sigma_n^{-1}\bm b_q)^2}\Big(-\frac{(\bm b_q^{\rm 5g})^H\bm h_{k,q}^{\rm 5g}\beta_{k,q}^{\rm 5g}x_{k,q}^{\rm 5g}}{\sigma^2} \notag \\
&\quad + \frac{(\bm b_{m,q}^\text{wf-d2d})^H\bm h_{j,q}^\text{wf-d2d}\beta_{j,q}^\text{wf-d2d}x_{k,q}}{\sigma_m^2}\Big) + \lambda_k(|\beta_{k,q}^{\rm 5g}x_{k,q}|^2-p_k^{\rm max}) \notag \\
&\quad - \omega_3T_q - 2\lambda_j\Big(\sum_{i\in{\cal N}_{j,m,q}^{\rm II}}(1-\alpha_{i,q})x_{i,q}\beta_{j,q}^\text{wf-d2d}\Big)x_{k,q}\beta_{j,q}^\text{wf-d2d} \notag \\
& \quad + \omega_2(C_{\rm 5g}-C_{\rm wf})=0,~~\forall k\in\cup_{j\in{\cal K}_{m,q}^\text{wf-d2d}}{\cal N}_{j,m,q} \\
&  \frac{2 w_1|\sum_{k=1}^Kx_{k,q}|^2}{(1+\bm b_q^H\bm \Sigma_n^{-1}\bm b_q)^2}\Big(-\frac{(\bm b^{\rm 5g}_{q})^H\bm h_{k,q}^{\rm 5g}\beta_{k,q}^{\rm 5g}x_{k,q}}{\sigma^2} \notag \\
&\quad +\frac{(\bm b^{\rm wf}_{m,q})^H \bm h_{k,q}^{\rm wf}\beta_{k,q}^{\rm wf}x_{k,q}}{\sigma_m^2} \Big) + \lambda_k(|\beta_{k,q}^{\rm 5g}x_{k,q}|^2 - |\beta_{k,q}^{\rm wf}x_{k,q}|^2 )\notag \\
& \quad  + \omega_2(C_{\rm 5g}-C_{\rm wf}) = 0,~\forall k\in {\cal K}_{m,q}^{\rm wf},
\end{align}
\end{subequations}
where $T_q\triangleq |{\cal P}_{k_0^{(q)}\rightarrow{\delta_{k_0^{(q)},q}}}|T_{\rm tx}$, $\forall q\in{\cal Q}$, and $m\in{\cal M}$.
\end{lemma}

 The obtained $\tilde{\alpha}^*_{k,q}$ in Lemma~\ref{lem.Relaxed-alpha} is not guaranteed to be an integer. In this case, we implement a rounding procedure to generate a feasible solution, $\alpha_{k,q}^*=[\tilde{\alpha}_{k,q}^*]\in\{0,1\}$, where $[x]$ denotes the nearest zero or one for $0\leq x \leq 1$.

\subsection{Updating Dual Variable Vector $\bm \lambda$}
Now that the RAT association and transmit optimization for a given $\bm \lambda$ have been obtained, problem (P2) can be solved via the dual problem \eqref{eq.prob-dual} by employing subgradient methods\cite{Boyd,Subgradient}. Specifically, at the $\ell$-th iteration, the dual function ${\cal D}(\bm \lambda)$ in \eqref{eq.dual-func} is evaluated with the given $\bm \lambda^{(\ell)}$. The dual variable vector $\bm \lambda^{(\ell)}$ are then updated via the subgradient method, i.e.,
\begin{align}\label{eq.dual-update}
 &   \lambda_k^{(\ell+1)} = \lambda_k^{(\ell)} + \gamma^{(\ell)}\left(\sum_{q=1}^Qp_{k,q}^{{\rm tx}}(\bm \alpha^{(\ell)}_q,\bm \beta_q^{(\ell)})-\tilde{p}_k\right),
\end{align}
where $\gamma^{(\ell)}$ is the diminishing step size at the $\ell$-th iteration. Note that the subgradient updates of \eqref{eq.dual-update} are guaranteed to converge when $\gamma^{(\ell)}$ is chosen to be sufficiently small.

\begin{algorithm}
\caption{for solving problem (P1) based on Lagrange Dual Decomposition and Primal Decomposition}
\begin{algorithmic}[1] 
\State {\bf Initialization:~} Given initial feasible dual variable vector $\bm \lambda^{(\ell)}$ for \eqref{eq.prob-dual}, the initial RAT selection variables $\{\bm \alpha_q^{(\ell)}\}$, and the prescribed accuracy $\epsilon$; set the initial iteration index as $\ell=0$.
\State {\bf Repeat:}
\begin{itemize}
\item Under the fixed dual variable vector $\bm \lambda^{(\ell)}$ and RAT selection vector $\bm \alpha_q^{(\ell)}$, obtain the WDs' transmit coefficients $\{\beta_{k,q}^{{\rm 5g}(\ell)},\beta_{k,q}^{{\rm wf}(\ell)},\beta_{k,q}^{{\rm wf-d2d}(\ell)}\}$ based on Lemma~\ref{lem.opt-beta};

\item Under the fixed dual variable vector $\bm \lambda^{(\ell)}$ and transmit coefficient vector $\bm \beta_q^{(\ell)}$, obtain the relaxed RAT selection solution of $\{\tilde{\alpha}^{(\ell+1)}_{k,q}\}$ based on Lemma~\ref{lem.Relaxed-alpha};

\item Obtain the binary RAT selection $\alpha^{(\ell+1)}_{k,q}=[\tilde{\alpha}^{(\ell+1)}_{k,q}]$ by rounding the continuous $\{\tilde{\alpha}^{(\ell+1)}_{k,q}\}$ to the nearest zero or one for $k\in{\cal K}$ and $q\in{\cal Q}$;

\item Obtain the updated dual variables $\bm \lambda^{(\ell+1)}$ based on \eqref{eq.dual-update};

\item Set $\ell\gets \ell+1$;
\end{itemize}

\State {\bf Until}
${\rm flag}(\ell) \triangleq \frac{|{\cal D}(\bm \lambda^{(\ell+1)})-{\cal D}(\bm \lambda^{(\ell)})|}{{\cal D}(\bm \lambda^{(\ell)})}<\epsilon$ satisfies the convergence condition.

\State {\bf Set}~$\alpha_{k,q}^{\star}\gets {\alpha}_{k,q}^{(\ell)}$, $\forall k\in{\cal K}$, $q\in{\cal Q}$;

\State {\bf Set}~$\bm \beta_{q}^{\star} \gets \bm \beta_{q}^{(\ell)}$, $\forall q\in{\cal Q}$;

\State {\bf Set}~ Obtain the optimal receive aggregation vectors $\{\bm v_q^{\star}\}$ with the updated $(\{\bm \alpha^{\star}_q\},\{\bm \beta^{\star}_q\})$ based on Lemma~\ref{lem.BF}.

\State {\bf Output:} Obtain $\{\alpha^\star_{q},\bm \beta^\star_q,\bm v_q^\star\}$ for problem (P1).
\end{algorithmic}
\end{algorithm}

\subsection{Summary and Complexity Analysis}

\begin{figure}
\centering
  \includegraphics[width=2.8in]{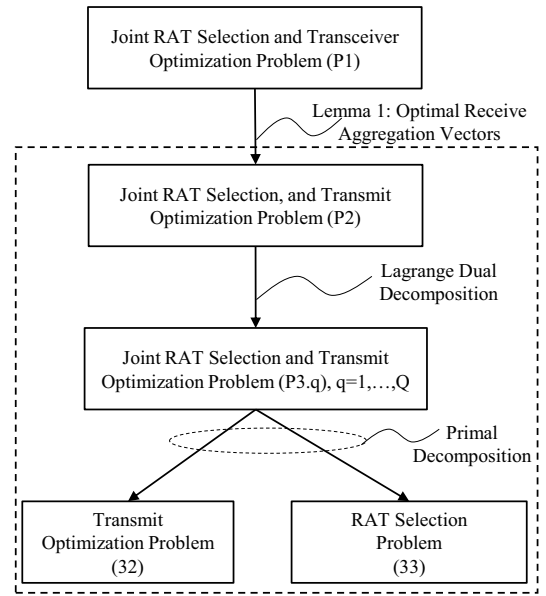}
  \caption{The relationship among the problems for the joint design of RAT selection and OTA aggregation of IVAs via 5G or WiFi for wireless MapReduce computation systems.}\label{fig:Relationship-Prob}
\end{figure}

 With $\{\bm \alpha_{q}^{\star},\bm \beta_{q}^{\star}\}$ obtained for problem (P2) based on the Lagrange duality method, we can obtain the optimal receive aggregation vectors $\{\bm v_q^{\rm 5g{\star}},\bm v_{m,q}^{\rm wf{\star}},\bm v_{m,q}^{\text{wf-d2d}{\star}}\}$ for problem (P1) based on Lemma~\ref{lem.BF}. Algorithm 1 is presented for obtaining the solution $\{\bm \alpha^{\star}_{q},\bm \beta^{\star}_q,\bm v^{\star}_q\}$ of problem (P1).

 The joint RAT selection and transceiver optimization problem (P1) is solved via a series of decomposition methods, as summarized in Fig.~\ref{fig:Relationship-Prob}. First, based on Lemma~\ref{lem.BF}, the optimal receive aggregation vectors $\{\bm v_q\}$ at the gNB/APs are obtained under the given RAT selection and transmit coefficients. By substituting the optimal receive aggregation vectors into problem (P1), we next obtain the transformed problem (P2), which is decomposed into the joint RAT selection and transmit optimization sub-problem (P3.$q$) for $q\in{\cal Q}$ via the Lagrange duality method. Based on the primal decomposition method, each problem (P3.$q$) for each $q\in{\cal Q}$ is further separated into the transmit optimization problem \eqref{eq.prob-Tx} and the RAT selection problem \eqref{eq.prob-RAT}.

\begin{figure}
\centering
  \includegraphics[width=3.5in]{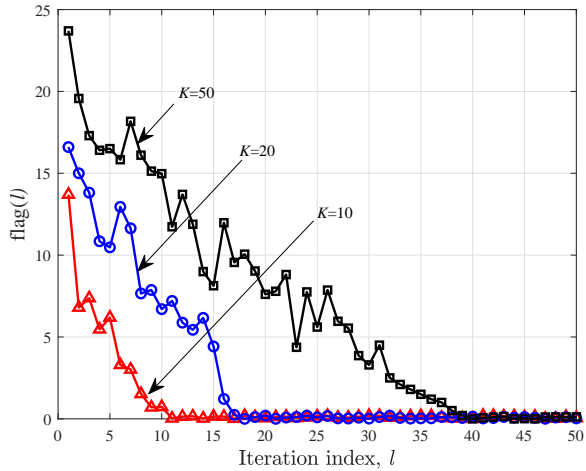}
  \caption{The convergence performance of proposed Algorithm 1, where the WiFi AP number is $M=4$.} \label{fig:Convergence}
\end{figure}

 Before leaving this section, we would like to note that the proposed Algorithm~1 (based on the Lagrange duality method and primal decomposition) for solving problem (P1) is guaranteed to converge to a stationary point. For Algorithm~1, the computational complexity is ${\cal O}((1+K+M)^{3.5}MQ)$ at each iteration, and the complexity of the outer Lagrangian multiplier update based on the subgradient method is a polynomial function of the dual problem dimension, i.e., $2K$ for ${\cal D}(\bm \lambda)$\cite{Boyd}. Towards achieving an optimal solution with an error tolerance $\epsilon$, it takes no more than $2K^{2.5}\log(RG/\epsilon)$ iterations for updating the dual variables, where $R$ denotes the distance between the initial point and the obtained stationary point and $G$ denotes the Lipschitz bound on the objective value of (P1)\cite{Scutari2017,Subgradient}. Therefore, the proposed Algorithm~1 only requires an affordable polynomial computational complexity of ${\cal O}((1+K+M)^{3.5}K^{2.5}MQ\log(RG/\epsilon))$ to find a solution for the joint RAT selection and transceiver design problem (P1) with a desired accuracy. The fast convergence of Algorithm~1 is corroborated by the numerical results in Fig.~\ref{fig:Convergence}, where $\rm{\rm flag}(\ell) \triangleq \frac{|{\cal D}(\bm \lambda^{(\ell+1)})-{\cal D}(\bm \lambda^{(\ell)})|}{{\cal D}(\bm \lambda^{(\ell)})}$, the error tolerance $\epsilon=10^{-3}$, the AP number $M=4$, and the remaining parameters are set to be the same as those in Section VI. It is observed in Fig.~\ref{fig:Convergence} that Algorithm~1 requires around 11, 18, and 40 iterations to converge into the desirable solutions under the cases with $K=10$, $K=20$, and $K=50$, respectively.

\section{Numerical Results}
In this section, we provide a series of simulations to show the effectiveness of our proposed joint RAT selection and transceiver design scheme for wireless multiuser MapReduce computing. In simulations, we consider a two-tier cell with radius of 500 meters~(m), where a gNB equipped with $N_{\rm 5g}=4$ antennas is in the center and $M$ APs, each equipped with $N_{\rm wf}=2$ antennas, are symmetrically placed around the circle with a radius of 200~m. The $K$ WDs are randomly and uniformly distributed in the cell. The pathloss between the gNB/APs and WDs is modelled as $30.6+37.6\log_{10}d$, where $d$ is the corresponding distance, and the standard derivation of shadow fading is 8 dB. The noise power at the gNB/AP receiver is set to be -174~dBW. The real and imaginary parts of each IVA are set to be uniformly distributed within the interval $[1,3]$, i.e., ${\rm Re}[{\rm IVA}_{q,n}]\sim {\cal U}[1,3]$ and ${\rm Im}[{\rm IVA}_{q,n}]\sim {\cal U}[1,3]$, $\forall q\in{\cal Q},n\in{\cal N}$. The number of data files and output functions are set to be $N=5000$ and $Q=200$, respectively. Regarding the map function computation at the WDs, the size of each data file is set to be $D=2000$ bits, the execution time is $T_{\rm map}=0.01$ seconds, and the number of CPU cycles to executing one bit is $C_k=1000$, $\zeta_k=10^{-28}$, $\forall k\in{\cal K}$. Unless specified otherwise, the energy budget for each WD $k\in{\cal K}$ is set to be $E_k=10$ Joules, the transmission time is $T_{\rm tx}=0.02$ seconds, and the error tolerance is $\epsilon=10^{-3}$ for the proposed Algorithm~1. The IVA transmission costs of 5G and WiFi access are set as 1 unit and 1/5 unit, respectively. We used a server with a Core i7-10710U, 1.6~GHz processor, and 16GB RAM to run all simulation-based experiments. All numerical results were obtained by averaging over 2000 independent realizations.

For performance comparison, we consider the following four baseline schemes.
\begin{itemize}
    \item{\em Only 5G Access Scheme for OTA Aggregation of IVAs:} In this scheme, only a single 5G gNB is deployed in the cell, where all the $K$ WDs are allowed to simultaneously transmit their aggregated IVAs to the gNB.
    \item{\em Only WiFi Access Scheme for OTA Aggregation of IVAs:} In this scheme, only $M$ WiFi APs are deployed in the cell, where all $K$ WDs are allowed to simultaneously transmit their aggregated IVAs to the WiFi APs.
    \item{\em Digital 16-QAM Transmission Scheme with 4-bit Quantization\cite{Luo06-baseline}:} In this scheme, each WD $k\in{\cal K}$ first employs a uniform quantizer to generate an unbiased 4-bit message for its aggregated IVA, and then adopts a 16-QAM (quadrature amplitude modulation) transmission scheme for the four quantized bits~\cite{Luo06-baseline}.
    \item{\em Branch-and-Bound (BnB) Optimal Scheme\cite{BnB2011}:} In this scheme, we employ the BnB approach \cite{BnB2011} to find the globally optimal solution for the original problem (P1). It serves as a lower bound for the existing schemes. The BnB scheme is proposed to optimally solve discrete and mixed-integer optimization problems via the state space search, but it generally has a very high computational complexity, especially when the WD number is large.
\end{itemize}

\subsection{Performance Tradeoff between MSE, Cost, and Delay}
\begin{figure}
\centering
  \includegraphics[width=3.5in]{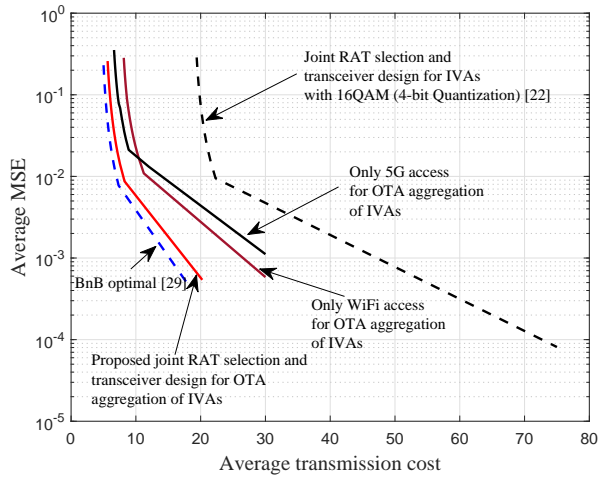}
  \caption{The average MSE performance versus the IVA transmission cost, where the WD number is $K=20$ and WiFi AP number is $M=4$.} \label{fig:MSE-Cost}
\end{figure}

\begin{figure}
\centering
  \includegraphics[width=3.5in]{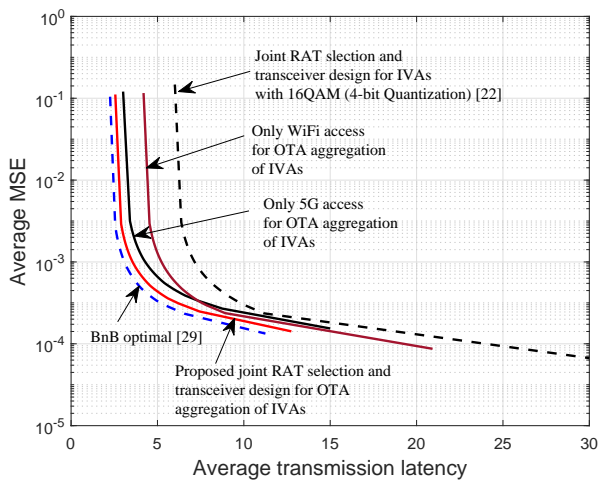}
  \caption{The average MSE performance versus the IVA transmission latency, where the WD number is $K=20$ and WiFi AP number is $M=4$.} \label{fig:MSE-Delay}
\end{figure}


Fig.~\ref{fig:MSE-Cost} first presents the tradeoff performance between the average MSE and the average IVA transmission cost for different schemes under the same latency conditions, where $K=20$ and $M=4$. The proposed joint RAT selection and transceiver design design scheme is observed to achieve a close MSE-cost tradeoff performance of the BnB optimal scheme. As compared with the only-5G-access and only-WiFi-access schemes, the proposed scheme achieves a significant gain on both MSE and cost performances. This implies the importance to exploit all the available multiple RATs in performing WDs' OTA aggregation of IVAs. By contrast, the digital joint RAT selection and transceiver design with 16-QAM transmission scheme performs inferior to the other schemes, which indicates the merit of analog OTA aggregation of IVAs over the digital counterpart. As expected, due to the lower transmission cost of WiFi access than that of 5G access, the baseline only-WiFi-access scheme is observed to achieve a substantial MSE gain at the large transmission cost regime, but it is not true at the small transmission cost regime under this setup.

Next, Fig.~\ref{fig:MSE-Delay} shows the tradeoff performance between the average MSE and the average IVA transmission latency for different schemes under the same transmission cost conditions. As in Fig.~\ref{fig:MSE-Cost}, the proposed scheme is observed to achieve closely to the lower bound of the BnB optimal scheme, and it outperforms the other three baseline schemes in both MSE and transmission latency performances. The baseline only-5G-access scheme outperforms the only-WiFi-access scheme at the small transmission latency regime, but it is not true as the transmission latency becomes larger. This is because the 5G access has a smaller latency than the WiFi access, since all the WDs have the 5G access and some WDs outside of the WiFi coverage have to implement the device-to-device communication to access a WiFi AP.

\subsection{Average MSE Performance}

\begin{figure}
\centering
  \includegraphics[width=3.5in]{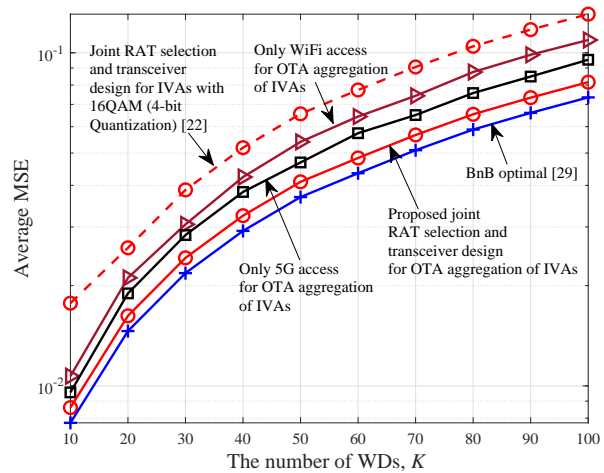}
  \caption{The average MSE performance versus the WD number $K$, where the WiFi AP number is $M=4$.} \label{fig:MSE-K}
\end{figure}

\begin{figure}
\centering
  \includegraphics[width=3.5in]{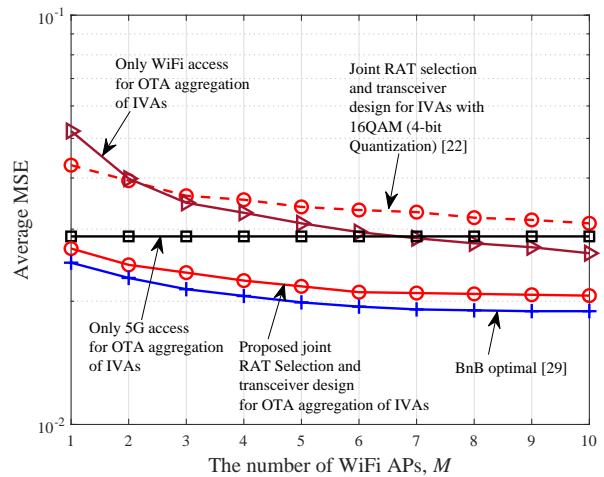}
  \caption{The average MSE performance versus the WiFi AP number $M$, where the WD number is $K=30$.} \label{fig:MSE-M}
\end{figure}

Fig.~\ref{fig:MSE-K} shows the average MSE versus the WD number $K$, where the WiFi AP number is $M=4$. For a fair comparison, all the schemes are evaluated under the same transmission cost and time latency conditions. The average MSE for all the schemes is observed to increase as the WD number $K$ increases. This is expected, since the IVA transmission with an increasing number of WDs implies a larger signal misalignment error. It is observed in Fig.~\ref{fig:MSE-K} that the proposed joint RAT selection and transceiver design scheme for OTA aggregation of IVAs achieves a close performance to that of the optimal BnB scheme, and outperforms the other baseline schemes. This illustrates the importance of exploiting both 5G and WiFi networks to reduce the computational MSE in analog OTA aggregation of IVAs. In addition, the baseline joint RAT selection and transceiver design with the 16-QAM scheme performs inferiorly to the only-5G-access and only-WiFi-access schemes for OTA aggregation, which implies the benefit of performing analog OTA aggregation of IVAs. Finally, the baseline only-5G-access scheme is observed to outperform the baseline only-WiFi-access scheme in this setup.

Fig.~\ref{fig:MSE-M} demonstrates the average computational MSE performance versus the WiFi AP number $M$, where $K=30$. Again, all the schemes are evaluated under the same transmission cost and time latency conditions for a fair comparison. Except for the baseline only-5G-access scheme, the average MSE of the other schemes decreases with the increasing of the WiFi AP number $M$, which indicates the benefit of deploying more WiFi APs in the cell for IVA transmission. In Fig.~\ref{fig:MSE-M}, it is again observed that proposed scheme achieves a close MSE performance to that of the BnB optimal scheme, and outperforms the other baseline schemes. This illustrates the effectiveness of tradeoff between the computational complexity and performance in the proposed scheme. Interestingly, there exists a threshold on the AP number $M$, as shown in Fig.~\ref{fig:MSE-M}; the baseline only-WiFi-access scheme performs inferiorly to the baseline only-5G-access scheme when $M$ is small (e.g., $m\leq 6$), but it is not true when the AP number $M$ becomes larger. The baseline 16-QAM scheme outperforms the baseline only-WiFi-access scheme when $M\leq 2$, but performs inferiorly to the other schemes as $M$ increases.


\subsection{Average Cost Performance}

\begin{figure}
\centering
  \includegraphics[width=3.5in]{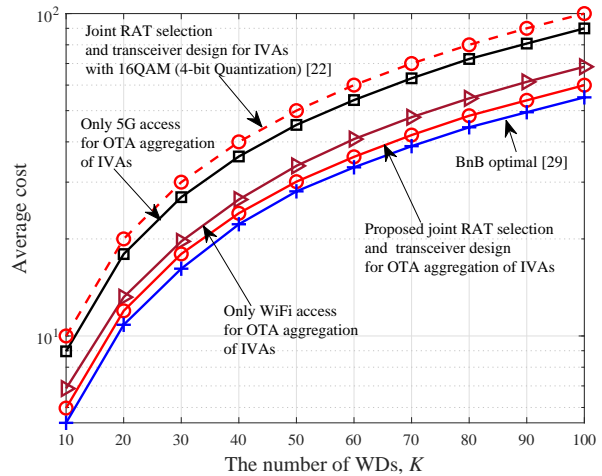}
  \caption{The average cost versus the WD number $K$, where the WiFi AP number is $M=4$.} \label{fig:Cost-K}
\end{figure}

\begin{figure}
\centering
  \includegraphics[width=3.5in]{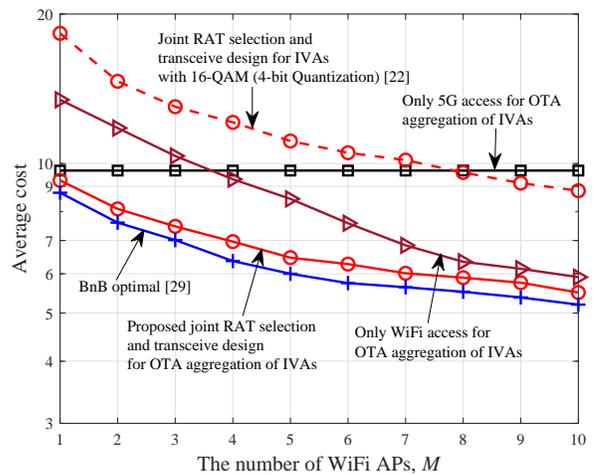}
  \caption{The average cost versus the WiFi AP number $M$, where the WD number is $K=20$.} \label{fig:Cost-M}
\end{figure}

In this subsection, we evaluate the average cost performance for different schemes. For a fair comparison, all the schemes are evaluated in Figs.~\ref{fig:Cost-K} and \ref{fig:Cost-M} to achieve the same computational MSE performance for OTA aggregation of IVAs.

Fig.~\ref{fig:Cost-K} shows the average IVA transmission cost versus the WD number $K$, where the WiFi AP number is $M=4$. The transmission cost of all the schemes is observed to increase with the increasing of the WD number $K$, which incurs an increasing amount of IVA transmission. In Fig.~\ref{fig:Cost-K}, the proposed joint design scheme achieves a slightly larger cost than the BnB optimal scheme, and outperforms the other baseline schemes. With a significantly smaller cost than the 5G access in IVA transmission, the baseline only-WiFi-access scheme is observed to outperform the only-5G-access scheme and the baseline digital 16-QAM transmission scheme. In addition, the baseline only-5G-access scheme outperforms the baseline digital 16-QAM transmission scheme, which further indicates the cost-saving benefit of exploiting analog OTA aggregation of IVAs.

 Fig.~\ref{fig:Cost-M} shows the average IVA transmission cost versus the WiFi AP number $M$, where the WD number is set as $K=20$. Apart from the only-5G-access scheme, the cost of the other schemes decreases as $M$ increases. Similar to Fig.~\ref{fig:Cost-K}, the baseline BnB optimal scheme is observed in Fig.~\ref{fig:Cost-M} to achieve the smallest cost among all the schemes. The proposed joint RAT selection and transceiver design scheme achieves a close performance to that of the BnB optimal scheme, and it outperforms the other three schemes. This implies the cost-saving benefit by simultaneously exploiting both the analog OTA aggregation of IVAs and multi-RAT access properties. The baseline only-WiFi-access scheme is observed to outperform the only-5G-access scheme at a large $M$ value (e.g., $M\geq 4$), but it is true when $M$ becomes smaller. This is because a larger $M$ value implies a larger number of WDs selecting APs for their IVA transmissions, thereby reducing the transmission cost. A similar phenomenon is observed for the digital IVA transmission scheme, when compared to the only-5G-access scheme.

\section{Conclusion}
 In this paper, we studied the tradeoff among the computational MSE, transmission cost, and time delay for OTA aggregation of IVAs in wireless multiuser multi-RAT MapReduce computing systems. Towards minimizing the weighted sum of the computational MSE, transmission cost, and transmission time delay, we jointly optimized the RAT selection and transceiver design for OTA aggregation of IVAs, subject to the energy budget constraints for Map function computing and IVA transmission per WD. A computational efficient algorithm was developed to optimize the WDs' RAT selection, the WDs' transmit coefficients for OTA aggregation of IVAs, and the receive aggregation vectors at the gNB/APs. Numerical results were provided to reveal the merit of the proposed joint multi-RAT OTA aggregation scheme, as compared with the existing single-RAT schemes and a digital transmission scheme.


\ifCLASSOPTIONcaptionsoff
  \newpage
\fi



\begin{thebibliography}{1}
\bibitem{Sar14}
S. Barbarossa, S. Sardellitti, and P. D. Lorenzo, ``Communicating while computing: Distributed mobile cloud computing over 5G heterogeneous networks,'' {\em IEEE Signal Process. Mag.}, vol. 31, no. 6, pp. 45--55, Nov. 2014.

\bibitem{Chiang16}
M. Chiang and T. Zhang, ``Fog and IoT: An overview of research opportunities,'' {\em IEEE Internet Things J.} vol. 3, no. 6, pp. 854--864, Dec. 2016.

\bibitem{JunZhang17}
Y. Mao, C. You, J. Zhang, K. Huang, and K. B. Letaief, ``A survey on mobile edge computing: The communication perspective,'' {\em IEEE Commun. Surveys Tuts.}, vol. 19, no. 4, pp. 2322--2358, 4th Quar. 2017.


\bibitem{Feng18}
F. Wang, J. Xu, X. Wang, and S. Cui, ``Joint offloading and computing optimization in wireless powered mobile-edge computing systems,'' {\em IEEE Trans. Wireless Commun.}, vol. 17, no. 3, pp. 1784--1797, Mar. 2018.

\bibitem{TCOM19}
F. Wang, J. Xu, and Z. Ding, ``Multi-antenna NOMA for computation offloading in multiuser mobile edge computing systems,'' {\em IEEE Trans. Commun.}, vol. 67, no. 3. pp. 2450--2463, Mar. 2019.

\bibitem{Bi2020}
J. Yan, S. Bi, Y. Zhang, and M. Tao, ``Optimal task offloading and resource allocation in mobile-edge computing with inter-user task dependency,'' {\em IEEE Trans. Wireless Commun.}, vol. 19, no. 1, pp. 235--250, Jan. 2020.


\bibitem{MapReduce-08}
J. Dean and S. Ghemawat, ``MapReduce: Simplified data processing on large clusters,'' {\em Commun. ACM }, vol. 51, no. 1, pp. 107--113, 2008.

\bibitem{MapReduce-10}
J. Dean and S. Ghemawatt, ``MapReduce: A flexible data processing tool,'' {\em Commun. ACM}, vol. 53, no. 1, pp. 72--77, 2010.

\bibitem{SongzeLi-18}
S. Li, M. A. Maddah-Ali, Q. Yu, and A. S. Avestimehr, ``A fundamental tradeoff between computation and communication in distributed computing,'' {\em IEEE Trans. Info. Theory}, vol. 64, no. 1, pp. 109--128, Jan. 2018.

\bibitem{SongzeLi-17}
S. Li, Q. Yu, M. A. Maddah-Ali, and A. S. Avestimehr, ``A scalable framework for wireless distributed computing,'' {\em IEEE/ACM Trans. Net.}, vol. 25, no. 5, pp. 2643--2654, Oct. 2017.

\bibitem{FanLi-19}
F. Li, J. Chen, and Z. Wang, ``Wireless MapReduce distributed computing,'' {\em IEEE Trans. Info. Theory}, vol. 65, no. 10, pp. 6101--6114, Oct. 2019.

\bibitem{Tao21}
F. Xu, S. Shao, and M. Tao, ``New results on the computation-communication tradeoff for heterogeneous coded distributed computing,'' {\em IEEE Trans. Commun.}, vol. 69, no. 4, pp. 2254--2270, Apr. 2021.

\bibitem{Shi-19}
K. Yang, Y. Shi, and Z. Ding, ``Data shuffling in wireless distributed computing via low-rank optimization,'' {\em IEEE Trans. Signal Process.}, vol. 67, no. 12, pp. 3087--3099, Jun. 2019.

\bibitem{Simeone-19}
S. Ha, J. Zhang, O. Simeone, and J. Kang, ``Wireless map-reduce distributed computing with full-duplex radios and imperfect CSI,'' in {\em Proc. IEEE SPAWC}, Cannes, France, Jul. 2019, pp. 1--5.

\bibitem{Feng21}
F. Wang and V. K. N. Lau, ``Multi-level over-the-air aggregation of mobile edge computing over D2D wireless networks,'' 2021. [Online]. Available: \url{https://arxiv.org/abs/2105.00471}

\bibitem{Boche2013}
M. Goldenbaum, H. Boche, and S. Stanczak, ``Harnessing interference for analog function computation in wireless sensor networks,'' {\em IEEE Trans. Signal Process.}, vol. 61, no. 20, pp. 4893--4906, Oct. 2013.

\bibitem{Boche2015}
M. Goldenbaum, H. Boche, and S. Stanczak, ``Nomographic functions: Efficient computation in clustered Gaussian sensor networks,'' {\em IEEE Trans. Wireless Commun.}, vol. 14, no. 4, pp. 2093--2105, Apr. 2015.

\bibitem{Chen18}
L. Chen, N. Zhao, Y. Chen, F. R. Yu, and G. Wei, ``Over-the-Air computation for IoT networks: Computing multiple functions with antenna arrays,'' {\em IEEE Internet Things J.}, vol. 5, no. 6, pp. 5296--5306, Dec. 2018.

\bibitem{Zhu19}
G. Zhu and K. Huang, ``MIMO over-the-air computation for high-mobility multimodal sensing,'' {\em IEEE Internet Things J.}, vol. 6, no. 4, pp. 6089--6103, Aug. 2019.

\bibitem{FengOTA20}
F. Wang and J. Xu, ``Optimized amplify-and-forward relaying for hierarchical over-the-air computation,'' in {\em Proc. IEEE GLOBECOM Workshops}, Dec. 2020, pp. 1--6.

\bibitem{Olga-15}
O. Galinina, A. Pyattaev, S. Andreev. M. Dohler, and Y. Koucheryavy, ``5G multi-RAT LTE-WiFi ultra-dense small cells: Performance dynamics, architecture, and trends,'' {\em IEEE J. Sel. Areas Commun.}, vol. 33, no. 6, pp. 1224--1240, Jun. 2015.

\bibitem{Luo06-baseline}
J.-J. Xiao, S. Cui, Z.-Q. Luo, and A. Goldsmith, ``Power scheduling of universal decentralized estimation in sensor networks,'' {\em IEEE Trans. Signal Process.}, vol. 54, no. 2, pp. 413--422, Feb. 2006.

\bibitem{YeLi-15}
G. Yu, Y. Jiang, L. Xu, and G. Y. Li, ``Multi-objective energy-efficient resource allocation for multi-RAT heterogeneous networks,'' {\em IEEE J. Sel. Areas Commun,}, vol. 33. no. 10, pp. 2118--2127, Oct. 2015.

\bibitem{XinWang-21}
B. Wu, T. Chen, K. Yang, and X. Wang, ``Edge-centric bandit learning for task-offloading allocation in multi-RAT heterogeneous networks,'' {\em IEEE Trans. Veh. Technol.}, vol. 70, no. 4, pp. 3702--3714, Apr. 2021.

\bibitem{Luo11}
Q. Shi, M. Razaviyayn, Z.-Q. Luo, and C. He, ``An iteratively weighted MMSE approach to distributed sum-utility maximization for a MIMO interfering broadcast channel,'' {\em IEEE Trans. Signal Process.}, vol. 59, no. 9, pp. 4331--4340, Sep. 2011.

\bibitem{Hao17}
Y. Hao, Q. Ni, H. Li, and S. Hou, ``On the energy and spectral efficiency tradeoff in massive MIMO-enabled HetNets with capacity-constrained backhaul links,'' vol. 65, no. 11, pp. 4720--4733, Nov. 2017.

\bibitem{Multi-Obj1}
R. T. Marler and J. S. Arora, ``Survey of multi-objective optimization methods for engineering,'' {\em Struct. Multidiscipl. Optim.}, vol. 26, no. 6, pp. 369--395, Apr. 2004.

\bibitem{Multi-Obj2}
Y. Sun, D. W. K. Ng, J. Zhu, and R. Schober, ``Multi-objective optimization for robust power efficient and secure full-duplex wireless communication systems,'' {\em IEEE Trans. Wireless Commun.}, vol. 15, no. 8, pp. 5511--5526, Aug. 2016.

\bibitem{BnB2011}
S. Boyd and J. Mattingley, ``Branch and bound methods,'' Dept. Elect. Eng., Stanford Univ., Stanford, CA, USA, Tech. Rep.,
May 2011. [Online]. Available: \url{https://stanford.edu/class/ee364b/lectures/bb_notes.pdf}

\bibitem{Scutari2017}
G. Scutari, F. Facchinei, and L. Lampariello, ``Parallel and distributed methods for constrained nonconvex optimization--Part I: Theory,'' {\em Trans. Signal Processing}, vol. 65, no. 8, pp. 1929--1944, Apr. 2017.

\bibitem{Burd96}
T. D. Burd and R. W. Brodersen, ``Processor design for portable systems,'' {\em J. VLSI Signal Process. Syst.}, vol. 13, nos. 2--3, pp. 203--221, 1996.

\bibitem{Goldsmith}
A. Goldsmith, {\em Wireless Communications}. Cambridge, U.K.: Cambridge Univ. Press, 2005.

\bibitem{Boyd}
S. Boyd and L. Vandenberghe, {\em Convex Optimization}. U.K.: Cambridge Univ. Press, 2004.




\bibitem{Kay93}
S. M. Kay, {\em Fundamentals of Statistical Signal Processing, Estimation/Detection Theory}. Englewood Cliffs, NJ: Prentice-Hall, 1993.

\bibitem{Subgradient}
S. Boyd, L. Xiao, and A. Mutapcic, ``Subgradient methods,'' Stanford Univ., Stanford, CA, USA, Lect. Notes EE392o, Oct. 2003. [Online]. Available: \url{https://web.stanford.edu/class/ee392o/subgrad_method.pdf}

\bibitem{Matrix-Inverse}
C. D. Meyer, {\em Matrix Analysis and Applied Linear Algebra}. Philadelphia, PA, USA: SIAM, 2000.

\end{thebibliography}
\end{document}